\documentclass[lettersize,journal]{IEEEtran}
\usepackage{amsmath,amsfonts}
\usepackage{algorithm}
\usepackage{array}
\usepackage{textcomp}
\usepackage{stfloats}
\usepackage{url}
\usepackage{verbatim}
\usepackage{graphicx}
\usepackage[justification=centering]{caption}
\usepackage{subfigure}
\usepackage{bm}
\usepackage{threeparttable}
\usepackage{booktabs}
 \usepackage[ natbib=true, style=numeric,sorting=none]{biblatex}
\begin{document}
\title{Is it a Real CD Mismatch in Interdomain Routing?}

\author{Letong Sun, Xingang Shi, Fengyan Han, Xia Yin, Zhiliang Wang, Han Zhang}

\maketitle
\begin{abstract}

In inter-domain routing, a packet is not always forwarded along the Autonomous System (AS) level path determined by the BGP routing protocol.
This is often called control-plane and data-plane (CD) mismatch, which allows for flexible traffic control, but also leads to operation and security issues. We systematically analyze this phenomenon with path pairs collected from 128 pairs of vantage points over more than 5 years, and use multiple IP-to-AS mapping methods to compare CD paths. What is interesting is that, working at such a large scale in turn helps us design a novel method to fairly evaluate the accuracy of various existing mapping methods, and further develop a new mapping method, i.e., LearnToCorrect, that can correct more than 70\% mapping errors of the state-of-the-art one. 
Then we devise to identify real mismatches with LearnToCorrect, and estimate that the real-mismatch ratio in the wild is typically less than 6\%. At last, we use our proposed methods to detect routing security issues, which are previously difficult to accurately find out. 

\end{abstract}

\section{Introduction}
The Internet is consisted of many Autonomous Systems (ASes) who use the Border Gateway Protocol (BGP) protocol to exchange routing information. Normally, a packet should be routed along an AS-level path determined by this control-plane protocol, as specified in BGP update messages or BGP routing tables. On the other hand, traceroute is the standard way to check how packets are actually forwarded along IP-level paths. It is known that, after translating the IP addresses in a traceroute path to the corresponding owner ASes, the resulting AS-level path might differ from that announced by BGP, a phenomenon called control-plane and data-plane mismatch (CD mismatch). 

CD mismatch is actually part of the design of the Internet. It allows for flexible traffic control, but also causes operation and security issues due to traffic detour, loop or black hole \cite{redirection}. Studying its prevalence and patterns helps a better understanding of our Internet, but faces several obstacles. 
First, IP-to-AS mapping, as an essential prerequisite for comparing the control-plane AS-level and data-plane IP-level paths, is nontrivial. Addresses can be shared or borrowed within Internet Exchange Points (IXPs), among sibling ASes of the same organization, or among neighboring ASes~\cite{ref_sig_mao, ref_hyun_incongruities, ref_jsac}. 
Although many mapping methods~\cite{ref_snmpv3,ref_info_mao,ref_router2as_assign,ref_bdrmap,ref_mapit,ref_bdrmapit,ref_hoiho2} have been proposed, due to a lack of ground-truth, little is known about their general accuracy and adequacy for our task. 
Second, estimating Internet-wide CD mismatch prevalence requires a substantial quantity of comparable control and data plane path pairs. Third, a few frequently occurring mismatched hops (or segments) can result in a large number of mismatched pairs, introducing bias into metrics such as trace-level or segment-level mismatch ratios. 
Previous studies on CD mismatch \cite{ref_sig_mao,ref_hyun_incongruities,ref_jsac,ref_tma} rely on traditional mapping methods, collect paths from a limited number of vantage points (VPs) within a short period, and use biased metrics which often overestimate the extent of mismatch. Their estimation results can be arguable (sometimes even contradictory), and the mismatch cases they find contain significant noises for analyzing routing issues.

In this paper, we first conduct a large scale comparison on CD paths to get a thorough view on the extent of CD mismatch, where multiple state-of-art mapping methods are experimented. We then use the comparison results to devise a method that can generate a large set of quasi-ground-truth, which is used to fairly evaluate different mapping methods, and further inspires a new mapping method that outperforms existing ones. At last, we use our refined methods to study mismatch cases that are related to routing security issues. 
Our contributions are as follows:



\begin{itemize}
\item We conduct a large-scale and long-term comparison on pairwise inter-domain control-plane and data-plane paths. 
In particular, we collect comparable traceroute and BGP records over 5 years from 128 pairs of dispersed VPs, 
and use 6 IP-to-AS mapping methods respectively to cross-check the 
Internet-wide mismatch ratios. The scale and depth of our analysis far exceed previous ones, which use data from at most 8 VP pairs in no more than a year, and give results under only one or two error-prone mapping methods. (\$III)

\item We propose the VISV algorithm to find a large set of accurate mappings from our CD comparison results. This set can be used as a qusi-ground-truth to fairly evaluate the accuracy of IP-to-AS mapping methods. It eliminates a major bias in typical large-scale traceroute records, where a few frequently occurring but wrongly mapped IPs may affect many traces. Our evaluation shows that BdrmapIt~\cite{ref_bdrmapit}, a state-of-the-art mapping method, outperforms traditional methods which rely on routing table or PeeringDB~\cite{ref_peeringdb} information,  
and it may be further improved with appropriate supplement data. (\S IV)

\item We design a novel mapping method named LearnToCorrect to facilitate our CD analysis. 
It constructs a learning model by correlating the VISV mapping results and the patterns in traceroute paths, 
and then utilizes this model to find and correct around 70\% wrong (and unaccomplished) mappings made by BdrmapIt. 
Although its training involves both contorl and data plane paths, 
the final model has a good spatial and temporal adaptability. 
On validation sets that are disjoint (i.e., collected from different VPs)  with path pairs used to train the model, it still achieves an average F1 score of 86\%. So LearnToCorrect can also be used as a general method to improve IP-to-AS mappings on traceroute paths. (\S V)

\item Based on the above methods, we devise to identify real-mismatch cases, and estimate that the real-mismatch ratio in the wild is typically less than 6\%. In particular, we exemplify on detecting complex routing security issues, e.g., hidden hijacks~\cite{ref_rov++} and bogus links~\cite{ref_mitm}, which are usually considered difficult to detect in reality. (\S VI)

\item We contribute our analysis code and results\footnote{https://github.com/cdmismatch/cdmismatch.} to the community to facilitate reproduction and future research.

\end{itemize}

\section{Background and related work}
Mismatch between control-plane path (computed by routing protocols) and data-plane path (how packets are actually forwarded) is allowed by design in the Internet to make traffic control more flexible, where network managers can redirect traffic by locally configured  policies. It can also occur in an undesired way due to intricacies and flaws in our distributed routing protocols. For example, iBGP may cause inconsistency between routing and forwarding within a domain~\cite{ref_forwarding_detours}, and false BGP messages may be utilized to manipulate inter-domain traffic~\cite{ref_rov++,ref_spurious_bgp}. The network community has recognized the resulting operation and security issues manifested in traffic detour, 
black hole, etc~\cite{redirection, ref_argus}. However, the actual prevalence of CD mismatch in the Internet has been studied only by very few researchers.

Mao et al.~\cite{ref_sig_mao} and Hyun et al.~\cite{ref_hyun_incongruities} are the first to measure inter-domain CD mismatch. They employ a small number of VPs (8 and 3 VPs respectively) which provide both traceroute and BGP data within a few months, translate an IP level trace into an AS level trace using active BGP routes in the routing table (which we call the RIB-match method), and assume that most mismatches are mapping errors. Their conclusion is that most mapping errors are caused by IXPs and sibling ASes. 
Fiore et al.~\cite{ref_tma} apply a similar method on data collected from 8 VPs from several days to a few months, and conclude that sibling ASes are the main causes of mapping errors. 
Zhang et al.~\cite{ref_jsac} point out that a vanilla comparison of CD path pairs may be biased due to a few IPs frequently occurring in a group of traceroute records. They use data collected from 4 VPs within a month and analyze mismatch at AS segment level. Instead of only using RIB-match mappings, they use a combination of RIB-match, the Internet Routing Registry data (IRR~\cite{ref_irr}) and DP-MIN (a mapping method discussed later) to perform IP-to-AS mappings, 
and conclude that neighboring ASes sharing an address block contribute greatly to mapping errors. 
Due to the limited measurement scale and unjustified mapping accuracy of existing studies, their conclusions on the ratio and cause of mismatches are often contradictory, and there lacks a systematic study on real mismatched cases.

As can be seen, accurate IP-to-AS mapping, i.e., telling an IP (especially that of a router interface) is used by which AS, is an essential step in CD mismatch analysis. 
A commonly used method, which we call RIB-match, is to find an IP's longest-matching prefix in a BGP routing table, i.e. BGP Routing Information Base (RIB), and use the AS originating that prefix as the mapping result. However, RIB-match is error-prone in practice, since an AS often uses addresses not owned/announced by itself. For example, two neighboring ASes have to use an address block owned by one of them on their inter-AS link, ASes peering at an IXP often use addresses provided by the IXP, and sibling ASes may borrow IPs from each other without explicitly announcing them in BGP. It is even possible that a same address block is announced by multiple ASes, either legitimately or maliciously. In addition, when a router receives a traceroute probe, it may respond with a third-party address (i.e., address of another router interface that neither receives nor forwards this probe), and this address may happen to be owned by another AS ~\cite{ref_tp1,ref_tp2,ref_tp3}. 

There are various efforts to achieve more accurate IP-to-AS mappings. A simple way is to use information published by network 
operators as a complement to RIB. For example, information about peering at IXPs or other co-locating facilities is often registered in PeeringDB~\cite{ref_peeringdb}, and many networks register their prefixes in IRR. However, such information is still incomplete, and may even be outdated.

Improving IP-to-AS mappings has also been considered in CD mismatch analysis. Mao et al.~\cite{ref_sig_mao} propose to modify RIB-match with some heuristics to make AS-level traceroute paths and BGP paths match as much as possible. They further design a dynamic programming algorithm~\cite{ref_info_mao}, which we call DP-min, to change mapping results and minimize the computed mismatch ratio. Although we agree with the idea that more accurate mappings result in lower mismatch ratios, solely minimizing mismatch ratios to determine the final mapping results is ill-designed. 
For example, 
when BGP route aggregation is performed by an AS $A$, the BGP path will terminate at $A$, and the corresponding data-plane path may pass through $A$ and further arrives at AS $B$, leading to a real CD mismatch. However, a brute force correction made by DP-min on the last hops would obscure the real destination AS $B$. Similarly, it may hide certain advanced routing hijacks where the data-plane path differs from the announced BGP path. 

Another family of solutions rely on typical patterns of a group of traceroute traces.  Bdrmap~\cite{ref_bdrmap}, MAP-IT~\cite{ref_mapit} and BdrmapIt~\cite{ref_bdrmapit} propose heuristics to identify inter-domain links first, and then infer the owner ASes of interfaces. In particular, BdrmapIt combines the other two, and votes among related interfaces appearing in different traces to determine the owner AS. Vrfinder~\cite{ref_vrfinder} further improves on inferring out-bound interfaces where BdrmapIt might fail, and it has been integrated into the latest BdrmapIt code~\cite{ref_bdrmapit_code}. 
These solutions can also collaborate with a technique called {\itshape IP alias resolution}~\cite{ref_midar,ref_iffinder,ref_rocketfuel,ref_radargun}, which can group different interface IPs into individual routers, and feed this information to BdrmapIt for voting. Midar~\cite{ref_midar} and Iffinder [16] are two representative alias resolution methods. They rely on the observation that interfaces on the same router often share a common IP ID counter, or a common interface is used for replying to probes on unused TCP/UDP ports. 
Kapar~\cite{ref_kapar} uses graph analysis to increase the number of grouped aliases at a cost of lower precision. 
These alias resolution methods have been used in CAIDA Macroscopic Internet Topology Data Kit (ITDK)~\cite{ref_itdk} 
to work with BdrmapIt \cite{ref_bdrmapit_code} (or a voting method \cite{ref_router2as_assign} based on RIB-match). Besides, Albakour et al.~\cite{ref_snmpv3, ref_bgp_alias} utilizes various protocol (e.g., SNMPv3, SSH and BGP) patterns as fingerprints to identify routers. Hoiho~\cite{ref_hoiho1} uses machine learning to compose regular expressions and extracts AS numbers from router hostnames. Its training uses the mapping results of BdrmapIt, and its results can be fed back into BdrmapIt. Although these advanced mapping methods seem to be promising, 
no large-scale and comparable evaluations have been performed on them, and they 
have not been used in studying CD mismatch.



\section{Compare control-plane and data-plane paths}\label{sec-comp}
\subsection{Data collection}

To get an in-depth and unbiased view of CD mismatch, we try to collect comparable control and data plane paths from as many geo-distributed vantage points (VPs) as possible. We rely on open platforms like RouteViews~\cite{ref_routeviews} and RIPE RIS~\cite{ref_riperis} to collect BGP routes from BGP routers, which we call control-plane VPs (C-VPs). There are around 4,260 C-VPs peering with the route collectors (RCs) in these two platforms, but most C-VPs only report customer routes to the RCs. 
In order to use RIB-match to find the longest matching prefix for a given IP, we choose a C-VP only when it reports more than 800K BGP routes to its RC, since currently a BGP routing table in the Default-Free-Zone contains around 1M entries. In the end, we can find 426 such C-VPs. 
On the other hand, we rely on CAIDA Ark~\cite{ref_ark} and RIPE Atlas~\cite{ref_atlas} to collect data-plane paths. Ark has around 200 routers, and on each day, around 10 of them conduct Paris traceroute~\cite{ref_paris_traceroute}, each to the whole IPv4 space (in /24 blocks). Atlas has over 12,000 probing hosts, and on each day, they collaborate to traceroute the whole routable IPv4 space ~\cite{ref_atlas_5051}, i.e., each host only covers around 100 /24 address blocks. We call these routers or hosts data-plane VPs (D-VPs).


The next step is to find (C-VP, D-VP) pairs who provide comparable paths. Ideally, we want that a C-VP and a D-VP record paths to the same destination in the same period, and they are in the same device so that the control plane path of C-VP can be regarded as the same as that of D-VP. We first group C-VPs and D-VPs according to the ASes they reside in. There are around 300 ASes where both C-VP and D-VP can be found on each day. Since in control plane, routers within an AS may adopt different routes to reach the same destination (e.g., when they are in different areas), we further use city-level geographical information to find pairs of C-VP and D-VP in the same AS and the same city. Such information for D-VPs are provided by the corresponding platforms, and for C-VPs, we check PeeringDB~\cite{ref_peeringdb} (if they peer with RC at IXPs) or use an inference algorithm \cite{ref_hoiho2} (if they have domain names) to exact their geo-location.
After this step, we find 5 (C-VP, Ark D-VP) and 123 (C-VP, Atlas D-VP) pairs. 

To study the long-term trend of CD mismatch without bringing excessive overhead, we use the BGP paths and traceroute paths recorded by these VPs on the 15th\footnote{Or the nearest date when both BGP and traceroute data are available.} of each month from 2018.01 to 2022.12 (this is the latest data we could get since Ark only publishes traceroute data older than a year). 
On each selected day, there are $36 \sim 128$ (C-VP, D-VP) pairs that have records. In total, our VP pairs cover 69 ASes that distributed in different topological and geographical locations. Table \ref{tab_network_types} lists their distribution according to different types of networks (the classification of networks is based on the algorithm in \cite{ref_flexsealing} and the data provided by PeeringDB~\cite{ref_peeringdb}).

The above approach to find a pair of co-located C-VP and D-VP is not without flaw. Their geo-location information may be inaccurate, or the colocated C-VP and D-VP may still have different control-plane routes, especially when their routing policies may change during our long period of study. To eliminate such defects and make our analysis more convincing, for a pair of control and data plane paths that are provided by a VP pair for the same destination at the same time, if the second-hop ASes in the two paths differ under every mapping method we use, we deliberately discard this pair. In this way, we 
eliminate potential errors introduced by improper selection of VPs. For interested readers, we also provide an analysis on these discarded paths in 
Appendix \ref{append-first-bifurcate}, 
where the above defects can be observed, and are confirmed by the network operators we contact.


We divide path pairs into per-VP-per-date datasets according to the corresponding VP pairs and date, so that we can study the trend of CD mismatch both in spatial and in temporal dimensions. 
However, since each Altas D-VP only cover a small number of prefixes on each day, for path pairs reported by (C-VP, Atlas D-VP) pairs, we only divide them into per-date datasets to avoid magnified variation. After this step, we have constructed 300 one-source-all-destinations (Ark) datasets, each of which contains around 760K path pairs on average, and 60 multi-source-mult-destination (Atlas) datasets, each of which contains around 10K path pairs on average. 


\begin{table}[!t]
\centering
\setlength\tabcolsep{5pt}
\caption{Distribution of D-VPs in different networks.}\label{tab_network_types}
\begin{tabular}{cccccc}
\hline
  &\emph{Tier-1}&\emph{large ISP}&\emph{small ISP}&\emph{stubs}&\begin{tabular}[c]{@{}c@{}}\emph{content}\\\emph{providers}\end{tabular}\\
\hline
Ark&0&1&4&0&0\\
Atlas&11&24&46&24&18\\
\hline
\end{tabular}
\end{table}

\subsection{Data preprocessing} \label{sec-cleans-data}
\textbf{Cleansing traceroute data.}
Although both Ark and Atlas use Paris traceroute \cite{ref_paris_traceroute} to avoid pitfalls of standard traceroute, anomalies caused by load balancing, spoofing~\cite{ref_loop}, etc., may still exist. 
We parse the data with Scamper~\cite{ref_scamper} and filter traces which Scamper flags as containing loops. 
We also filter traces where multiple responding IPs occur at the same hop. This often occurs when 
probes in the same traceroute procedure are forwarded along different paths. 
At last, we filter traces whose destination cannot match any prefix in the corresponding C-VP's routing table. However, we keep traces that contain unresponsive hops (represented by *) and those that have not reached the final destination, since such cases are non-negligible and contain useful information.  

Table ~\ref{tab_traceroute_ab} lists the percentage of these imperfect traceroute paths, where 4.6\% traces contain loops (\emph{loop}), 1.4\% have multiple IPs responding at the same hop (\emph{mul-resp}), 2.1\% do not have corresponding BGP prefix (\emph{no-BGP}). Over 60\% traces have unresponsive hops (\emph{unresp}) and 52\% traces do not reach the intended destinations (\emph{incomplete}).

\begin{table}[!t]
\centering
\setlength\tabcolsep{5pt}
\caption{Percentage of imperfect traceroute paths.}\label{tab_traceroute_ab}
\begin{tabular}{ccccc}
\hline
  \emph{loop}&\emph{mul-resp}&\emph{no-BGP}&\emph{unresp}&\emph{incomplete}\\
\hline
4.6\%&1.4\%&2.1\%&60.3\%&52.2\%\\
\hline
\end{tabular}
\end{table}

\begin{table}[!t]
\centering
\setlength\tabcolsep{5pt}
\caption{Percentage of special BGP paths.}\label{tab_bgp_ab}
\begin{tabular}{ccccc}
\hline
\emph{loop}&\emph{priv-ASN}&\emph{AS-SET}&\emph{dup-ASN}&\emph{IXP}\\
\hline
0.1\%&0.0\%&0.0\%&15.7\%&2.3\%\\
\hline
\end{tabular}
\end{table}

\textbf{Cleansing BGP data.}
BGP paths also have anomalies that need to be filtered. First, some BGP paths have loops, often caused by BGP poisoning~\cite{ref_flexsealing}. Second, private AS numbers can be erroneously embedded. Third, route aggregation introduces AS sets that cannot be translated back to paths. Besides filtering BGP paths with these anomalies, from the remaining paths, we also remove duplicated ASes, 
as well as IXP ASes that operate IXP route servers \footnote{Such IXP ASes do not participate in data-plane forwarding.}. Table~\ref{tab_bgp_ab} lists the percentage of each type of those BGP paths. We can see that loops (\emph{loop}), AS sets (\emph{AS-SET}) and private AS numbers (\emph{priv-ASN}) rarely occur, while 15\% BGP paths contain duplicated ASes (\emph{dup-ASN}) and 2\% contain IXP ASes (\emph{IXP}).

\subsection{IP-to-AS mapping methods} \label{sec-mapping-methods}
Besides the simple RIB-match algorithm, we also use several advanced mapping methods introduced in Sec \S II
. This is to expose the negative impact on CD mismatch analysis introduced by a single mapping method, 
while also offering a comparison on the accuracy of different mapping methods. Details of the comparison will be discussed in $\S$\ref{sec-cmp-methodology}. 


As analyzed before, the DP-min method~\cite{ref_info_mao} adjusts its mapping results according to the matching ratio, and thus it hurts the capability to find real mismatches. So we do not include it in our experiments. In total, we use 6 mapping methods, i.e., RIB-match, RIB+PeeringDB, BdrmapIt, Midar\&Iffinder(MI)+BdrmapIt, Fingerprint+BdrmapIt and Hoiho+BdrmapIt. 
In RIB+PeeringDB, we complement RIB with the data registered in PeeringDB~\cite{ref_peeringdb}. 
For BdrmapIt, we run its latest code~\cite{ref_bdrmapit_code} on the traceroute data in each dataset. We take special care on time consistency, making sure that all the data required by BdrmapIt\footnote{Data needed by BdrmapIt includes RIR data\cite{ref_rir}, AS-to-org mappings~\cite{ref_asorgs}, AS-relationships and customer-cones~\cite{ref_asrel}, PeeringDB data~\cite{ref_peeringdb}, etc.} is of the same date as (or closest to) our dataset. Actually we have tested BdrmapIt on one dataset, several datasets within a day or even within a year, and its mapping results seldom conflict.

For MI+BdrmapIt, we can only get MI's IP-to-Router and MI+BdrmapIt's IP-to-AS mappings generated from the Ark measurement data on a few days (which are provided by CAIDA ITDK~\cite{ref_itdk}, respectively in 2018.03, 2019.01, 2019.04, 2020.01, 2020.08, 2021.03 and 2022.02). So we apply the method only to the CD comparison on the Ark datasets. For datasets without the corresponding mappings on exactly the same time, we tried in two ways. The first is to use the available MI IP-to-Router mappings on the closest day, feed them into BdrmapIt to generate IP-to-AS mappings. We think this is reasonable since IP-to-Router mappings typically do not vary too much in a short period. We also tried directly using MI+BdrmapIt IP-to-AS mappings provided by Ark on those days, filling the gap in a way similar to how we fill IP-to-Router mappings. However the latter method gives us a less satisfying results in both trace-level and IP-level metrics. In particular, on datasets far from the days when MI+BdrmapIt data is available, up to 20\% hops are left as ``unmap'', and we see a clear the-farther-the-worse effect. So we only show results of the former in this paper. 

For Hoiho+BdrmapIt, we first use the hostnames and regular expressions provided by Hoiho ~\cite{ref_hoiho_data} to generate IP-to-AS mappings. Note that for better accuracy, we only use the regexes which Hoiho labels as \emph{good} and \emph{promising}. We then feed Hoiho's results into BdrmapIt for refinement (setting the parameter \emph{as-hints} in BdrmapIt to be true). Since Hoiho uses the alias resolution generated by MI, and its data is available only for 4 days (respectively in 2018.03, 2019.01, 2019.04 and 2020.01), we apply Hoiho only to the Ark datasets, and fill the gap with available Hoiho data on the closest date.

For Fingerprint+BdrmapIt, we use the alias resolution data built on SNMPv3, SSH and BGP fingerprints~\cite{ref_snmp_data,ref_bgp_alias}, and feed it into BdrmapIt. Since this alias resolution data is calculated in 2023.10, we apply Fingerprint+BdrmapIt only to our latest datasets (i.e., 2022.12). 

\begin{figure}
  \centering
  \includegraphics[width=.95\linewidth]{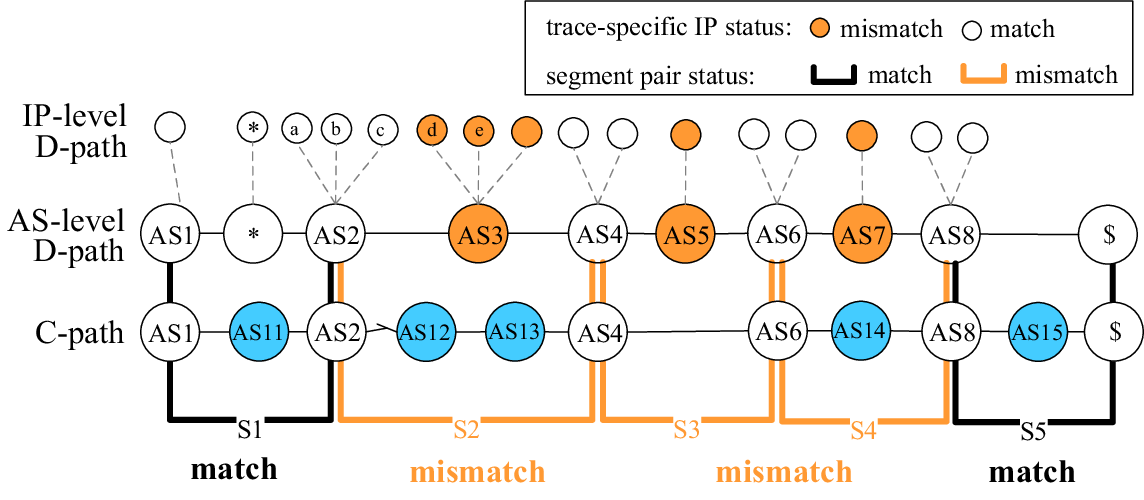}
  \caption{Break paths into segments in CD comparison.} \label{fig_path_segs_with_ips}
\end{figure}

\begin{figure*}[!tb]
      \subfigure[Group I.]{
      \begin{minipage}[t][0.44\width]{0.44\linewidth}%
          \includegraphics[clip,width=\textwidth]{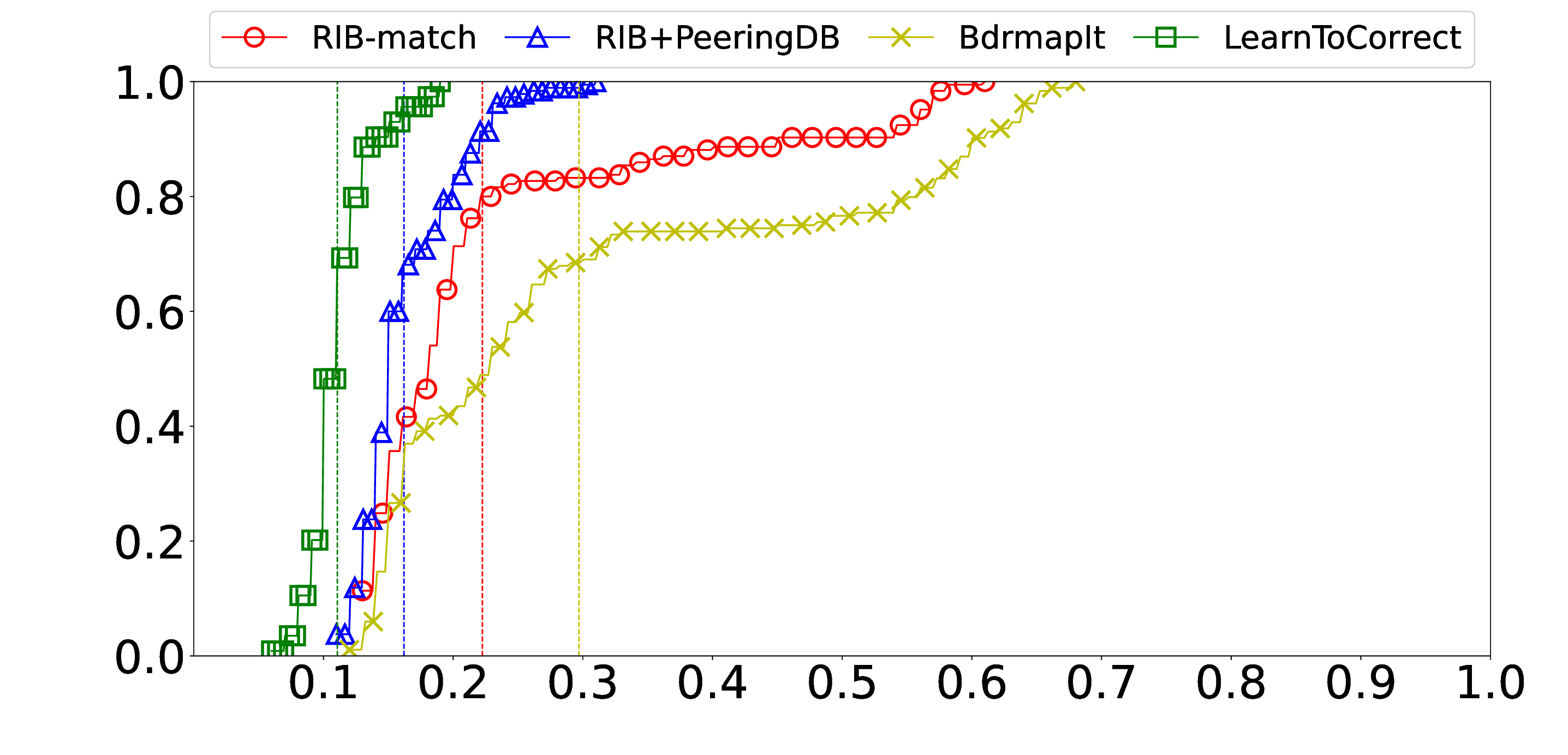}
      \end{minipage}
      }\label{fig_trace_level_rate_base}
      \subfigure[Group II.]{
      \begin{minipage}[t][0.44\width]{0.44\linewidth}%
          \includegraphics[clip,width=\textwidth]{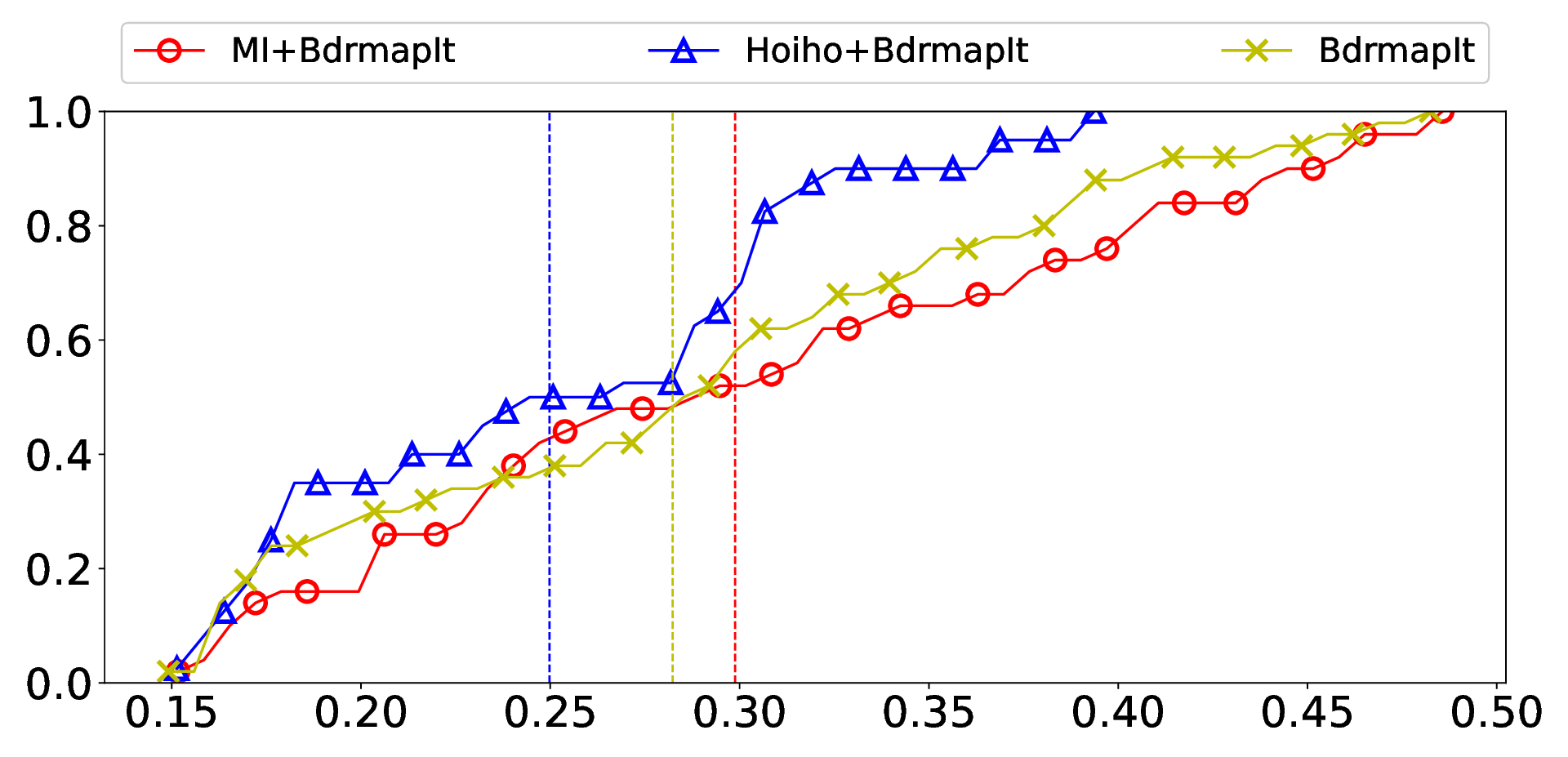}
      \end{minipage}
      }\label{fig_trace_level_rate_midar}
  \caption{CDF of trace-level {\itshape mismatch} ratio under different mapping algorithms.}
  \label{fig_trace_level_mismatch}
\end{figure*}

\subsection{Path pair comparison result} \label{sec-cmp-methodology}
For an IP-level data-plane path (D-path) reported by traceroute for a destination $d$, we find its longest-matching prefix in the corresponding C-VP's RIB or update messages, 
and use the BGP path with the closest timestamp as the control-plane path (C-path). 
To compare such a pair of CD paths, 
we first translate the latter into AS-level using each specific mapping method. 
Unresponsive hops, unmapped IPs (i.e., when the mapping method fails) and IPs mapped to an IXP AS are left as *, and will be treated as wildcards in comparison. 
If multiple consecutive IPs are mapped to the same AS, we only keep one AS. 
Although we have cleansed data as described in \S\ref{sec-cleans-data}, the comparison is not as simple as comparing two strings. 
We have to deal with various deficiencies in D-paths, including ``widecard'' hops, early terminations, etc.

For a pair of C-path and D-path, we first append an ending mark (e.g., "\$") to each one, then break them further into segment pairs, such that (1) two consecutive segments in the same path have one common endpoint, and (2) in each segment pair, the AS-level traceroute path segment (i.e. D-segment) and the BGP path segment (i.e. C-segment) must match at both ends, but have no overlap in the middle. In this way, we break the two paths into comparable segments as short as possible. Fig.\ref{fig_path_segs_with_ips} depicts a toy example, where the two paths 
are broken into five segment pairs (i.e., $S1$ to $S5$). 
A segment pair is labeled as {\itshape match} if both the D-segment and the C-segment has no middle hops, or if the D-segment has only wildcard middle hops and is no shorter than the C-segment (e.g., $S1$), or if it is the last segment but has no middle hops (e.g., $S5$  due to early termination of traceroute). 
Otherwise, the segment pair will be labled as {\itshape mismatch} (e.g., $S2$, $S3$ and $S4$). 
We regard a path pair as \emph{mismatch} if any of its segment pair is labelled as \emph{mismatch}. 
Otherwise, the path pair is regarded as \emph{match}.

We perform the above comparison on our datasets under each of the mapping methods introduced in \S\ref{sec-mapping-methods}.  
As mentioned before, some mapping methods can only be applied to a part of the datasets due to its data availability, so we divide the six methods into three groups for a fair evaluation. Group I includes RIB-match, RIB+PeeringDB and BdrmapIt, where the mapping data is available for all our datasets. In $\S$\ref{sec-newmapping}, we will design a new mapping method called LearnToCorrect which can also be applied for all our datasets, so we add it to Group I for comparison.
Group II includes MI+BdrmapIt, Hoiho+BdrmapIt and BdrmapIt. We can only get MI and Hoiho's data for the Ark datasets, so we compare them only in that scope, with BdrmapIt as a baseline. Group III includes Fingerprint+BdrmapIt and BdrmapIt, where we compare them only on datasets on 2022.12.

Fig.~\ref{fig_trace_level_mismatch} shows the comparison results for the first two groups, 
where the CDF of the trace-level {\itshape mismatch} ratios computed on each of our 360 datasets (across hundreds of VPs over 5 years) under each mapping method is depicted. For each CDF, we also plot a vertical line to indicate the corresponding mean value. In group I, we can see that RIB+PeeringDB (blue triangle) gives an average mismatch ratio of 17\%, which is lower than RIB-match (23\% in red circle) and BdrmapIt (27\% in yellow cross). 
In Group II, 
The average trace-level \emph{mismatch} ratio given by Hoiho+BdrmapIt (blue triangles) is 3\% lower than that given by BdrmapIt (yellow cross), and MI+BdrmapIt (red circle) gets the highest average mismatch ratios. Note that the average mismatch ratio under BdrmapIt in Group I (30\%) is higher than that in Group II (28\%), since BdrmapIt performs worse on the Ark datasets than on the Atlas datasets. This again demonstrates the necessity of comparing different mapping algorithms on the same dataset. Since Fingerprint+BdrmapIt and BdrmapIt give nearly identical results for the datasets in Group III, we omit the depiction. 

Overall, more than 75\% path pairs match under all mapping methods, which conforms to the common belief that 
\emph{CD paths match with each other in most time}~\cite{ref_hyun_incongruities, ref_jsac, ref_sig_mao, ref_info_mao}.
However, the mismatch ratios given by different mapping methods, and even those by BdrmapIt itself across different spaces and time, have non-negligible differences. 
In addition, a lower trace-level mismatch ratio does not necessarily suggest that one mapping method is more effective. For example, when an algorithm cannot find a mapping for an IP (which we name as \emph{unmap} hops)\footnote{There are various reasons a method cannot map out an IP. For example, RIB-match cannot map out IPs not announced by any AS (e.g., some ASes choose not to announce the networks used by their backbone routers); and BdrmapIt chooses not to map out an IP if it cannot get a determined result during the mapping process.}, we treat it as wildcard and do not trigger a mismatch in the comparison. A close look shows that, under RIB-match, there are 6\% \emph{unmap} IPs with each appearing in up to 60 D-paths on average, compared with the 3\% \emph{unmap} IPs under BdrmapIt with each appearing in only 28 D-paths. More importantly, the occurring frequency of wrongly mapped IPs can greatly impact the trace-level mismatch ratios.
All these raise the questions that how to fairly evaluate the mapping methods (\S\ref{sec-eval}), 
and how to design one with better accuracy (\S\ref{sec-newmapping}) to facilitate the analysis on mismatch cases (\S\ref{sec_study_mm_cases}).


\section{Evaluate Mapping Methods}\label{sec-eval}
To evaluate a mapping method, it is straightforward to check its accuracy if we have mapping ground truth, but achieving such data is non-trivial. 
As far as we know, ground-truth mappings used in earlier works are of a small diversity (e.g., $\sim$3K IPs in only 4 networks \cite{ref_bdrmapit}), and are not publicly available or reproducible. To solve this problem, we first devise to dig ground truth out of the traceroute paths. We couple that data with the private ones provided by network operators, and use them to compute the accuracy of each mapping method (\S\ref{sec_eval_IP_gt}). 
However, these two datasets are not large enough, and the evaluations then can not provide a good explanation on our earlier CD comparison results. We further design a method to fully utilize the CD comparison results to generate a rich set of IP-to-AS mappings that we have very high confidence (\S\ref{sec_visv}). We evaluate mapping methods by using these mappings as semi-ground-truth, and the result can very well explain our earlier findings on CD comparison (\S\ref{sec_eval_visv}). 


\subsection{Achieve ground truth from traceroute paths and by manual contact}\label{sec_eval_IP_gt}

When launching traceroute from a VP to a destination, if the VP and the destination are within the same AS,
it's very likely that the outcome traceroute path is an intra-AS path (i.e., it never goes out of this AS),  and all the IPs within belong to this AS. 
With this in mind, we check all traceroute paths provided by Ark and Atlas (during the same period as those used in our CD comparison), 
and select intra-AS paths where the traceroute destination is announced by the same AS at the same time as the starting point.
Note that we exclude paths where the prefix is announced by multiple ASes to avoid possible errors. 
Overall, we can validate more than 3000 intra-AS IPs from 19 ASes\footnote{On average, we have 513 Intra-AS IPs to validate daily mapping results.}. Most of theses IPs are from the Ark traces, since an Atlas probe only traceroutes to a small number of /24 prefixes, and the chance to find usable intra-AS paths is small.



\begin{table}
    \centering
    \begin{tabular}{cccc}
        \hline
        &Intra-AS&ISP&VISV\\
        \hline
        total number of IPs&3,029&457&879,571\\
        total number of ASes&19&2&38,815\\
        \hline
        RIB-match&2.1\%&26.2\%&10.0\%\\
        RIB+PeeingDB&2.1\%&26.2\%&9.8\%\\
        BdrmapIt (1.1\%--3.7\%~\cite{ref_bdrmapit}) &1.9\%&1.4\%&4.0\%\\
        MI+BdrmapIt&1.7\%&2.9\%&4.1\%\\
        Hoiho+BdrmapIt&1.6\%&1.3\%&3.9\%\\
        Fingerprint+BdrmapIt&1.8\%&1.4\%&3.2\%\\
        LearnToCorrect&0.5\%&0.8\%&1.4\%\\
        \hline
        VISV&0.1\%&0.0\%&--\\
        \hline
    \end{tabular}
    \caption{Average mapping error ratios with 3 validation sets.}
    \label{tab_eval_gt}
\end{table}

The Intra-AS column in Table \ref{tab_eval_gt} shows the average mapping error ratios of 8 mapping methods evaluated by the Intra-AS IPs, where VISV and LearnToCorrect will be introduced later. 
We can see that BdrmapIt achieves a slightely lower error ratio (1.9\%) than RIB-match or RIB+PeeringDB method (2.1\%). 
Feeding complement information (e.g., MI, Hoiho or SNMPv3/SSH/BGP alias resolution data) to BdrmapIt further reduces the errors slightly. 

We also contact two national-scale ISPs for manual validation. In the mapping results on our latest traceroute datasets (i.e., on 2022.12), we select an IP if at least one mapping method maps it to one of the two networks. 
In total, we can respectively validate 385 and 72 IPs in the two networks, and the validation results are shown in the ISP column of Table \ref{tab_eval_gt}. The error ratios of the two RIB based methods now surge from 2.1\% to 26.2\%, while BdrmapIt performs even better (1.4\%) than using the Intra-AS IPs. This divergence in results attribute to the mapping behavior of RIB-match on inter-AS link addresses as well as the IP positions of the two validation sets in our traceroute datasets. When a router receives traceroute probes, it normally responds with the IP on its inbound interface. This IP, when on one side of a provider-customer inter-AS link, is often allocated by the provider network. Consequently, if a traceroute passes through an inter-AS link from a provider network router to a customer network router, RIB-match is prone to map the inbound interface of the customer router to its provider network. The Intra-AS IPs are in the same AS as Ark or Atlas D-VPs that launch traceroute, so they often locate nearer to the source D-VPs and on a path segment from customer to provider. On the contrary, the two ISPs don't have D-VPs and their IPs often appear nearer to the traceroute destinations and on a path segment from provider to customer, where RIB-match makes more mistakes. RIB+PeeringDB only modifies RIB-match with records from PeeringDB, so it cannot fix this kind of flaws. On the other hand, BdrmapIt dedicates to correctly mapping the inter-AS link addresses, so it can noticeably reduce such errors.

On both the validation sets, BdrmapIt (with or without supplement information) has a better mapping accuracy than the RIB based methods. Considering that a wrong mapping is more likely to make a matching pair mismatch rather than turning a mismatching pair to match, this result contradicts with our earlier CD path comparison findings (Fig. \ref{fig_trace_level_mismatch} in \S\ref{sec-cmp-methodology}), where BdrmapIt gives obviously higher trace-level mismatch ratios than RIB-match and RIB+PeeringDB. 
Unfortunately, the two validation sets are too limited for us to explore the reason. A larger scale of accurate mapping data is needed to facilitate further investigations.

\subsection{Achieve semi-ground truth from CD comparison}\label{sec_visv}
Given the assumption that CD pairs usually match, massive paired CD paths can help us to get more accurate mappings, compared with most mapping methods that mainly rely on traceroute \emph{or} BGP data. Some studies~\cite{ref_sig_mao, ref_info_mao} have utilized this feature and tried to modify the RIB-match mapping results to make most path pairs match. However, their accuracy is affected mainly in two aspects. First, an IP can have multiple mapping candidates that make equal path pairs match. For example in Fig.\ref{fig_path_segs_with_ips}, IP $a$ is initially mapped to $AS2$. If this mapping matches all the related C-segments, and in all these C-segments $AS2$ follows $AS11$, then $AS11$ can also be a mapping for $a$ that matches all the C-segments. The occurrence of multiple optimal candidates becomes particularly prevalent when an IP appears in only a few segment patterns, such as IPs nearer to destinations appear in only one traceroute path. In such scenarios, previous methods either keep the original mapping or randomly pick a candidate, introducing unavoidable errors. Second, these methods do not consider the existence of real-mismatched path pairs, which, though proportionally small, can contribute to non-negligible inaccurate results.

\begin{figure}
  \centering
  \includegraphics[width=.9\linewidth]{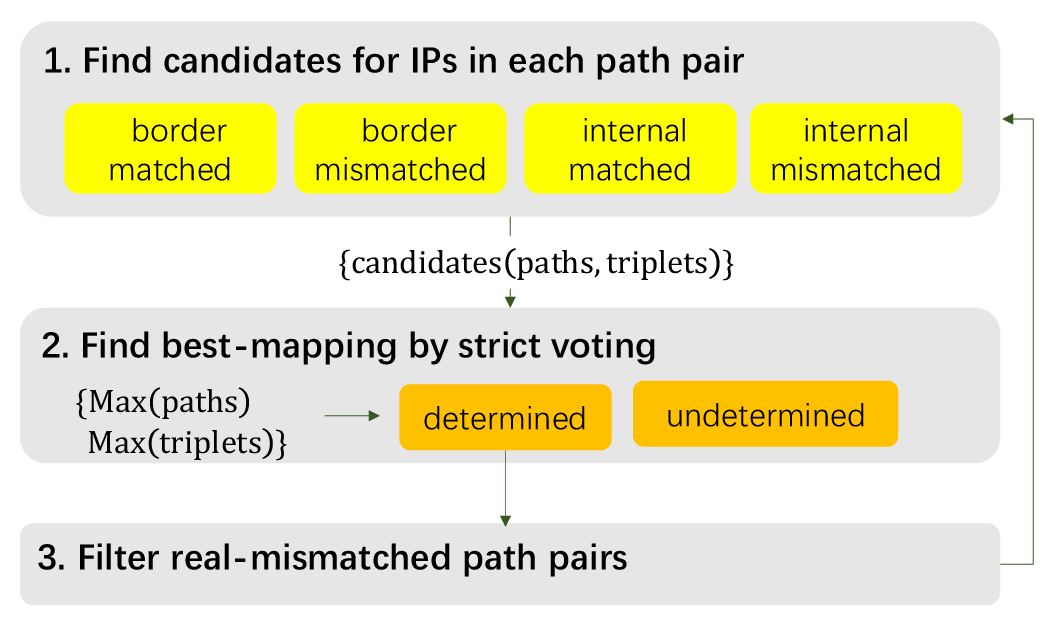}
  \caption{Workflow of VISV.} \label{fig_visv}
\end{figure}

To address the above challenges, we introduce Verdict by Iterative Strict Voting (VISV), an iterative approach designed to infer accurate mappings to the fullest extent possible. The workflow of VISV is depicted in Fig.\ref{fig_visv}. In each iteration, VISV first finds mapping candidates for an IP that can match the C-path in each path pair. This step is based on CD comparison described in \S\ref{sec-cmp-methodology}, where an example is illustrated by Fig.\ref{fig_path_segs_with_ips}.
When applying a mapping method, we label an IP as {\itshape unmap} if the algorithm cannot determine a result for this IP. For an IP that can be mapped to an AS, if the corresponding AS is a middle point (represented by orange circle) of a segment, 
we label this IP as {\itshape mismatch}, and otherwise (i.e., if the AS is an endpoint of a segment) as {\itshape match}.
We say a {\itshape match} or {\itshape mismatch} IP is \emph{internal}, if it is mapped to the same AS as both its preceding and succeeding IPs (ignoring unresponsive or \emph{unmap} hops). Otherwise, we say it is a \emph{border} IP. Candidate mappings for an IP are then collected based on the category it belongs to.
For a \emph{border match} IP (e.g., $a$ in Fig.\ref{fig_path_segs_with_ips}), besides the AS it is mapped to (i.e., AS2), 
the corresponding neighboring AS (i.e., AS11) in the C-path is also considered as a candidate. 
But for an \emph{internal match} IP, the result given by the mapping algorithm is the only candidate.
For a \emph{mismatch} IP (e.g., $d$), all the ASes in the corresponding C-segment (i.e., AS2, AS12, AS13, and AS4) 
are included as candidate mappings, and if it is \emph{internal} (e.g., $e$), the mapping result (i.e., AS3)
is also included as a possible candidate. 
The number of candidates actually reflects the uncertainty of the result given the current circumstance.
For a reference, Table \ref{tab_visv_candidate} lists the corresponding candidate mappings for the IPs $a$--$e$
in Fig.\ref{fig_path_segs_with_ips}.

To mitigate bias introduced by frequently occurring segment patterns across multiple path pairs, VISV also records the IP-level $predecessor-current-successor$ triples when collecting candidates for an IP. Then, the optimal candidate(s) for an IP is expected to make most path pairs as well as most distinct triples match. 
VISV designates the mapping status of an IP as \emph{determined} if it has only one such optimal candidate, otherwise \emph{undetermined}, signifying that an accurate result cannot be determined in the current iteration. 

The last step in an iteration is to check if any path pairs become mismatch given the \emph{determined} results. VISV counts these path pairs as real-mismatch and remove them from the path pair datasets for the next turn. 

VISV iteratively infers \emph{determined} mappings and filters out real-mismatched pairs. The process terminates when no more \emph{determined} mappings can be inferred.

\begin{table}
\centering
\setlength\tabcolsep{5pt}
\caption{Candidate mappings for hops in Fig.\ref{fig_path_segs_with_ips}.}\label{tab_visv_candidate}
\begin{tabular}{ccc}
\hline
 IP hop&category& candidate mappings\\
\hline
$a$&\emph{border match}&{$AS2$, $AS11$}\\
$b$&\emph{internal match}&{$AS2$}\\
$c$&\emph{border match}&{$AS2$, $AS12$}\\
$d$&\emph{border mismatch}&{$AS2$, $AS12$, $AS13$, $AS4$}\\
$e$&\emph{internal mismatch}&{$AS2$, $AS12$, $AS13$, $AS4$, $AS3$}\\
\hline
\end{tabular}
\end{table}

\begin{table}
\centering
\setlength\tabcolsep{5pt}
\caption{ \emph{determined} mappings (diagonal), common IPs with \emph{determined} mappings (upper triangle) and conflicting ratios (lower triangle) in primitive sets.}\label{tab_visv_consistency}
\begin{threeparttable}
\begin{tabular}{ccccccc}
\hline
&M1&M2&M3&M4&M5&M6\\
\hline
M1&\textbf{76,326}&74,858&66,740&66,251&66,458&66,806\\
M2&0.4\%&\textbf{80,443}&69,939&69,202&69,147&69,980\\
M3&0.1\%&0.7\%&\textbf{98,354}&91,315&92,589&97,255\\
M4&0.1\%&0.7\%&0.2\%&\textbf{96,037}&95,635&90,915\\
M5&0.1\%&0.7\%&0.2\%&0.1\%&\textbf{96,748}&91,026\\
M6&0.1\%&0.7\%&0.0\%&0.2\%&0.2\%&\textbf{98,249}\\
\hline
\end{tabular}
\begin{tablenotes}
    \item
   \item M1: RIB-match; M2: RIB+PeeingDB; M3: BdrmapIt;
    \item M4: MI+BdrmapIt; M5: Hoiho+BdrmapIt; M6: Fingerprint+BdrmapIt
\end{tablenotes}
\end{threeparttable}
\end{table}

\begin{figure}          
    \includegraphics[clip,width=\linewidth]{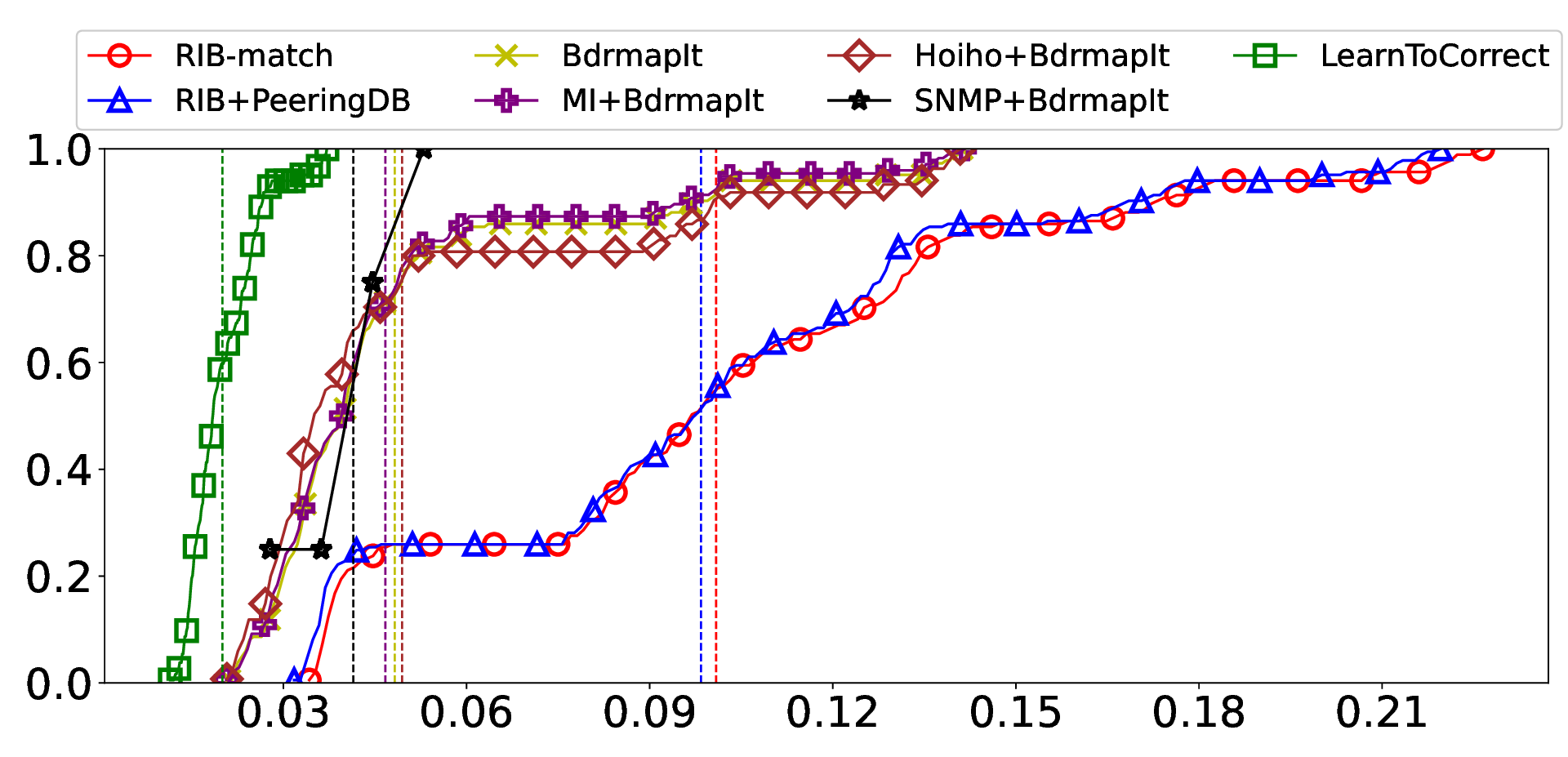}
  \caption{CDF of mapping error ratios of different mapping algorithms evaluated by VISV.}
  \label{fig_eval_visv}
\end{figure}



Since VISV needs a mapping method to start the CD comparison, we run VISV separately with each of the 6 mapping methods. 
In Table \ref{tab_visv_consistency}, we list statistics of the primitive sets computed based on each mapping method. We note that VISV is performed on each path pair dataset. The diagonal numbers are the numbers of \emph{determined} mappings in each set, averaged on the corresponding path pair datasets. A number in the upper triangle corresponds to the intersection of the sets computed by two mapping methods (i.e., common IPs that have \emph{determined} mappings), while a number in the lower triangle corresponds to the ratio of their conflicting mapping results. 


There are typically 220K intermediate IPs in each path pair dataset, excluding those appearing only as traceroute destinations. 
For 39\%--47\% of them, VISV can find a \emph{determined} mapping. 
VISV also achieves highly consistent results based on different mapping methods. There are 87\%$\sim$99\% common IPs with \emph{determined} mappings among either two mapping methods, and the ratios of conflicting results are less than 1\%. To further increase confidence, we select a \emph{determined} mapping as a final result only when it appears in at least two primitive sets, but does not conflict with results in any other primitive set. 
Overall, for all our 296 path pair datasets, VISV can determine mappings for 880K IPs, distributed in around 39K Ases. 

We validate VISV with the ground truth in \S\ref{sec_eval_IP_gt}, where VISV can determine mappings for 80\% of the intra-AS IPs and half of the ISP IPs. For these determined ones, the error ratio (last row in Table \ref{tab_eval_gt}) is only 0.1\% for the former, 
and no error occurs on the latter\footnote{VISV uses a simple voting scheme while it catches the main attributes such as path pairs, triplet patterns and real-mismatched cases. Other schemes are possible, but the results already show very high accuracy and provide a starting point for other more sophisticated rules.}. 
With such a high accuracy, we believe the results of VISV can be used as semi-ground truth to support a large scale evaluation on other mapping methods. 



\subsection{Evaluate mapping methods with VISV}\label{sec_eval_visv}


%

Fig.\ref{fig_eval_visv} depicts the CDF of mapping error ratios on our 360 path pair datasets for each mapping method. 
We can see that, RIB-match (red circle) and RIB+PeeringDB (blue triangle) both have error ratios around 10\%, with the latter slightly lower. Their average error ratios are also listed in the last column of Table \ref{tab_eval_gt}, which lie between the ratios estimated by the two ground-truth sets. As we have explained, RIB based methods perform worse when mapping the customer-side addresses of inter-AS links (ISP IPs) than mapping the provider-side addresses (Intra-AS IPs), and VISV provides evaluation data on both scenarios.
 

BdrmapIt based methods achieves much lower error ratios, with an average of 4\%. MI+BdrmapIt has nearly identical results as BdrmapIt, which conforms to the analysis in earlier work \cite{ref_bdrmapit}.
Supplementing Hoiho data (available for around 4\% IPs\footnote{With the method discussed in \S\ref{sec-mapping-methods}, we can generate about 10K Hoiho IP-to-AS mappings on each day.}) to BdrmapIt achieves improvement on dates when such data were collected, 
but the accuracy drops when the time is not consistent. 
Among the 36.7M (IPv4) IPs in the published router fingerprint alias resolution data~\cite{ref_bgp_alias}, we find around 48K of them appearing in our datasets and can be grouped to routers. However, many of these IPs are already correctly mapped by BdrmapIt, so supplementing the data (Fingerprint+BdrmapIt) does not increase the mapping accuracy significantly. 


\begin{table}[!t]
\centering
\setlength\tabcolsep{5pt}
\caption{Avg. IP freq., aggregated by evaluated results.}\label{tab_ip_span}
\begin{tabular}{cccc}
\hline
 &{\itshape correct}&{\itshape wrong}\\
\hline
RIB-match &28&17\\
RIB+PeeringDB &30&13\\
BdrmapIt  &29&\textbf{70}\\
MI+BdrmapIt  &28&\textbf{68}\\
Hoiho+BdrmapIt  &27&\textbf{43}\\
Fingerprint+BdrmapIt  &29&\textbf{70}\\
\hline
\end{tabular}
\end{table}

The mapping error ratios evaluated by VISV, again show a contradiction with the trace-level mismatch ratio results (i.e., BdrmapIt series have lower mapping error ratios but higher trace-level mismatch ratios). Given the large scale of VISV data, we can now look into the evaluated IPs and check their occurring frequencies (the number of paths an IP appears in) in each dataset. We find that, under each mapping method, the average IP occurring frequencies according to different evaluation results (\emph{right} or \emph{wrong})
differ. As shown in Table \ref{tab_ip_span}, a wrongly-mapped IP under BdrmapIt, MI+BdrmapIt or Fingerprint+BdrmapIt appears in 70 D-paths on average, which is much higher than that under RIB-match or RIB+PeeringDB ($\leq$ 30 D-paths). 
We then look into the datasets where the trace-level mismatch ratios under BdrmapIt are particularly higher ($\geq$ 50\%). We find that a large amount of mismatched path pairs are caused by a few wrongly-mapped IPs near the D-VPs. Neither MI+BdrmapIt nor Fingerprint+BdrmapIt correct these ``critical'' errors, while Hoiho+ BdrmapIt does, so it has lower trace-level \emph{mismatch} ratios compared with BdrmapIt (Fig.\ref{fig_trace_level_mismatch}). 

Based on the above results, we use BdrmapIt, which has low mapping error ratios and wide applications on all the datasets,
for our subsequent studies. For preparation, we check the causes of BdrmapIt's mapping errors. If an wrongly mapped IP is an IXP IP or of an IXP prefix (checked with PeeringDB), or if the operator AS is an IXP AS, we count the mapping error as caused by IXP. We find 9.4\% mapping errors fall in this category. For the other IPs, we check the relationship between the BdrmapIt mapping and the VISV mapping, and find that 11.6\% cases are siblings (using AS-to-organization data~\cite{ref_sibling_pam}), 71.8\% cases are neighbors (using AS relationship information \cite{ref_asrel}), and 7.2\% cases with unknown relationship. 

\section{An advanced mapping method}\label{sec-newmapping}
In this section, we introduce a method to identify and correct mapping errors in BdrmapIt. To make the method more widely used, we carefully devise not to rely on the path pair data which is usually hard to get, but only use traceroute paths and publicly available control-plane data as BdrmapIt did, rendering it a general mapping method.

\subsection{Identify mapping errors}
Our solution begins with a discovery that an wrong mapping often results in many of its occurring paths become irrational. For example, a wrongly mapped hop can introduce non-existing links in the AS-level D-paths or cause the paths to violate the valley-free principle. As undetected links and AS paths violating valley-free exist in reality, there is no direct relationship between an IP's mapping status (i.e., \emph{right} or \emph{wrong}) and the rationality of the paths it appears in. However, based on massive traceroute paths, we construct a machine-learning model to formalize the relationships between the mapping status of IPs (evaluated by the VISV data), and mapping features within the occurring paths. After being trained, this model can be used to identify mapping errors for all the IPs in the traceroute paths. 


\textbf{Construct the model.} In the modeling, it is essential to extract features that can capture the impacts of an IP's mapping on the paths. Suppose an IP $x$ is mapped to AS $S$ under BdrmapIt, we consider both its own attributes and its relationships with neighbors. 

First, as analyzed in \$\ref{sec_eval_visv}, 9.4\% of the mapping errors in BdrmapIt evaluated by VISV relate to IXP IPs or IXP ASes. So we check if $x$ or $A$ is IXP-related (using data from PeeringDB). Moreover, since $x$ can be operated by AS $A$ which announces it (the rationale of RIB-match), or by $A$'s neighbor (in inter-AS link address or third-party addresses), or by sibling or customer (in address reuse or address reallocation cases), or by $A$'s other peer (through IXP), we check the relationship between $S$ and $A$, considering that $S$ is prone to be wrong if it has no relationship with $A$. 

Second, as to the relationships with neighbors, we check respectively from the IP level and the AS level. On the IP level, we focus on if $S$ aligns with the mappings of either $x$'s predecessor or successor IPs. Specifically, we record the number of distinct IP prefixes (according to the announcements at the time) that are also mapped to $S$ in each traceroute path. We choose distinct prefixes instead of distinct IPs to avoid the impact of a shared prefix among neighbors or siblings, where multiple IPs within could be incorrectly mapped to a same AS. Continuous IP hops from different prefixes being mapped to $S$ indicates the high possibility that packets are traversed through $S$ in that path. 

On the AS level, we focus on the condensed AS-level D-path (i.e., without duplicate ASes) and check to see if $S$ has relationships with its predecessor and successor ASes. Furthermore, we check if the AS-level triples (with $S$ in the middle) are valley-free. In particular, we consider two triple patterns: two consecutive peer-to-peer (p2p) links and a provider-customer (p2c) link followed by a p2p
link. The two patterns, though violating valley-free, are often found in reality (according to BGP paths collected from the collector $RouteViews2$ on 2023.12.01 00:00, there are over 20\% routes violating valley-free, and 83\% of them belong to the above two patterns). We record these kinds of triples as \emph{part-valley-free}, as opposed to other \emph{valley-free} or \emph{non-valley-free} triples. 

To deal with the last hops in incomplete traceroute paths, we append an unresponsive hop ``*'' and the destination to the path end so that we can treat the original last hop as normal middle hops. At last, since unresponsive or unmapped hops are ignored in the previous steps, we record if there are such hops between $x$ and its IP neighbors, as well as $S$ and its AS neighbors, to consider their impact on the final result. 

To eliminate biases introduced by frequently occurring IPs, we record both relative ratios (e.g., the percentage of distinct \emph{valley-free} triples out of all the distinct triples an IP appears in) and absolute ratios (e.g., the percentage of \emph{valley-free} triples out of all the paths an IP appears in), on all the above relationship-concerned features. The exact set of features can be found in Appendix \ref{appendix_ml}.

Having extracted path features for the IPs that VISV covers, we split the BdrmapIt mapping status (evaluated by VISV data) into training and testing set.
We use XGBoost~\cite{10.1145/2939672.2939785} to train a binary-classification model based on the training set.
Specifically, the model assigns a probability score (ranging from 0 to 1) indicating the likelihood that a mapping is right. The model predicates the mapping as right if the score is greater than 0.5, and wrong otherwise. At last, we evaluate the model's performances using the testing set.

\textbf{Evaluate the model.} The effectiveness of the model include two aspects: (1) its ability to identify wrong mappings; (2) its adaptability when trained on some datasets and applied to others. For the first aspects, since in BdrmapIt the right mappings and wrong mappings are highly imbalanced (with an average ratio of 24:1 evaluated by VISV), we redefine "positive" samples as wrong mappings and "negative" samples as right mappings, in order to facilitate the evaluation using conventional metrics. In this context, precision ($TP/(TP+FP)$) refers to the fraction of correctly identified wrong mappings among all identified wrong mappings; 
recall ($TP/(TP+FN)$) refers to the fraction of correctly identified wrong mappings among all actual wrong mappings;
F-measure ($2*(PR*RC)/(PR+RC)$) provides a balance between precision and recall; and specificity ($TN/(TN+FP)$) refers to the fraction of correctly identified right mappings among all actual right mappings. 

For the second aspect, we divide the BdrmapIt mapping status data in three ways. First, we use the mapping status data on 2022.12 (the latest date available), randomly selecting 80\% of the data as training set and the remaining 20\% as testing set (A in Table \ref{tab-model}). This step evaluates the overall performance of the model on a set of traceroute paths. Second, we use the mapping status data from two Ark VPs on 2022.12 as training set, and the data from the other Ark VPs as well as Atlas VPs on 2022.12 as testing set. This step evaluates the model's adaptability to traceroute paths of different VPs on the same date—referred to as spatial adaptability. Since there are three Ark datasets and one Atlas dataset on 2022.12, we construct three different groups of training set and testing set (B$\sim$D in Table \ref{tab-model}). Third, we use all the mapping status data on 2022.12 as the training set, and data on earlier dates as testing sets, to test the model's temporal adaptability.

\begin{table}[]
    \centering
    \begin{tabular}{ccccccc}
    \hline
       &Training set&Testing set  & PR & RC & F1 & SPC\\
    \hline
       A&80\% of all&20\% of all &94\%&84\%&89\%&99\%\\
       B&\emph{ams-nl}+\emph{nrt-jp}&\emph{sao-br}+Atl &90\%&79\%&84\%&96\%\\
       C&\emph{nrt-jp}+\emph{sao-br}&\emph{ams-nl}+Atl &89\%&82\%&85\%&98\%\\
       D&\emph{sao-br}+\emph{ams-nl}&\emph{nrt-jp}+Atl &92\%&80\%&85\%&99\%\\
    \hline
    \end{tabular}
    \caption{Evaluate the model using different training and testing sets.}
    \label{tab-model}
\end{table}

Table \ref{tab-model} lists the results of the first four groups described before. In group $A$, the model achieves a precision (PR) of 94\%, recall (RC) of 84\%, F-measure (F1) of 89\%, and specificity (SPC) of 99\%. Although wrong mappings are extremely few in the whole dataset, our model not only identifies most wrong mappings (precison\textgreater 94\%), but also maintains a low rate of false positives (recall\textgreater 84\%). The spatial adaptability test in the other three groups also achieve high precision (around 90\%) and recall (around 80\%). Given that the D-VPs in the training and testing sets are geographically and topologically dispersed\footnote{The Ark D-VPs \emph{ams-nl}, \emph{nrt-jp} and \emph{sao-br} are operated by three small ISPs (with less than 50 customers) respectively in Europe, Asia and South America. Atlas D-VPs are operated by ASes from Tier-1 to stubs all over the world.}, the results demonstrate the good spacial adaptability of the model. 

\begin{figure*}
      \subfigure[Statistics of the results.]{
      \begin{minipage}[t][0.375\width]{0.375\linewidth}%
        \includegraphics[width=.9\linewidth]{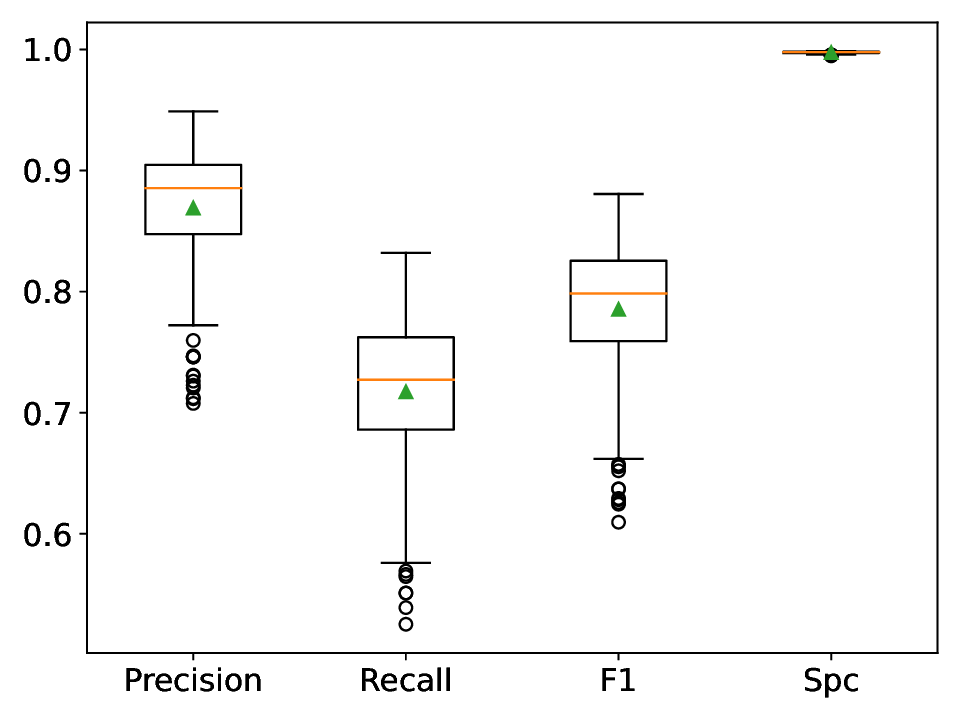}
      \end{minipage}
      }
      \subfigure[PR and RC along the timeline.]{
      \begin{minipage}[t][0.5\width]{0.5\linewidth}%
          \includegraphics[clip,width=\textwidth]{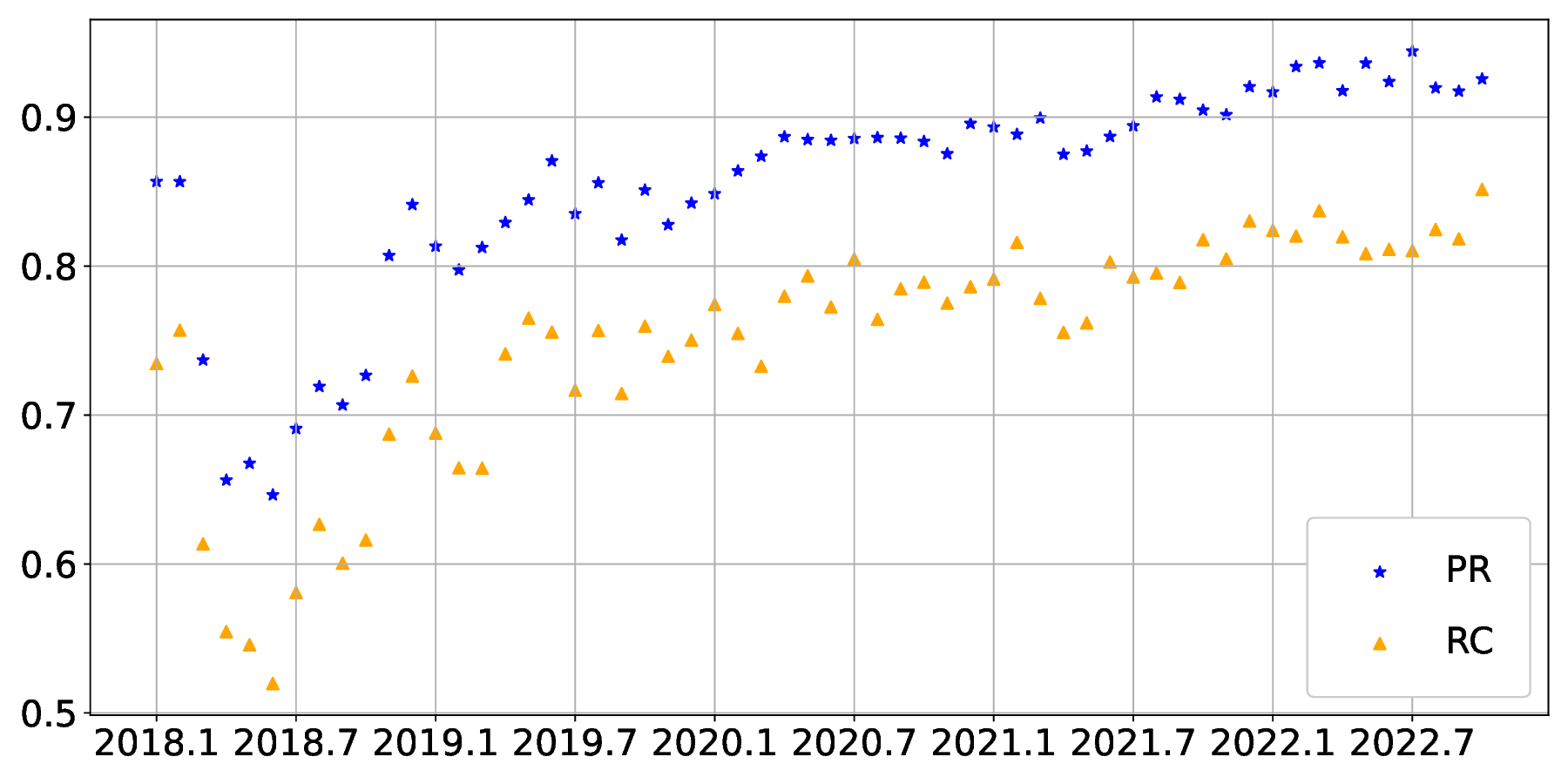}
      \end{minipage}
      }
  \caption{Results of the temporal adaptability test.}
  \label{fig-model-metrics-all}
\end{figure*}

Fig.\ref{fig-model-metrics-all}(a) depicts the statistics of the results in the temporal adaptability test, including the minimum, maximum, mean, median and 90\% confidence level of the results. We can see that the values of precision and recall exhibit significant variations, ranging in 70\%$\sim$94\% and 56\%$\sim$84\% respectively. F1-score shows the similar variability. Only SPC consistently exceeds 95\%, primarily owing to the large amount of the right mappings. Fig.\ref{fig-model-metrics-all}(b) further illustrates the temporal dynamics of precision (PR) and recall (RC). Both PR and RC exhibit a noticeable uptrend over time. Considering that the model is trained by datasets on 2022.12, it suggests that the forwarding behavior in the data-plane evolves along the time, which affects the performances of the model. However, PR has remained over 90\% since 2021.7, and RC has remained over 80\% since 2021.10, indicating that the trained model is effective  on datasets within at least a year. We can easily overcome the temporal limitation by periodic retraining using traceroute paths from available VPs, given the proved spatial adaptability of the model.

\subsection{Correct mapping errors}
Once the model is constructed, we try to correct mappings which is identified as wrong, without any further help of the path pair data. 
As analyzed in \$\ref{sec_eval_visv}, in 92.8\% of the wrong mappings, the corresponding right mapping (i.e., the VISV mapping) is either an IXP (which can be found in PeeringDB), or a neighbor in the traceroute paths, or has some AS relationship with the wrong mapping. Inspired by this, for a wrong mapping, we collect possible mapping candidates, select a most promising one, and replace it together with other wrong mappings in a systematic way.

We collect potential candidates in four ways: the mapping result of RIB+PeeringDB; the AS-level neighbors in the traceroute paths; the siblings/providers/customers of the wrong mapping result; and the ASes that, when combined with the AS-level predecessors and successors, can form valley-free triples. According to the proportions of the mapping error causes (\$\ref{sec_eval_visv}), we group the candidates in the above order.

To select a best candidate for an IP, we first put each candidate as the IP's current mapping into the model and get a score. However, simply choosing the highest score can be error-prone, especially when some candidates have nearer scores. For such candidates (i.e., whose score differences are less than a threshold $D$\footnote{We set $D$ to 0.1 and 0.2 respectively in the correcting process, and the final results show little difference.}), we select the candidate in the highest-ranked group first, and then choose the one that makes the most triples valley-free in case of a tie.


A more challenging issue is that an wrongly mapped IP can severely distort the scores of its neighbors, since the relationship with neighbors impact many important features in the learning model. 
A wrong mapping would either make a correctly-mapped neighbor being ill-scored, or make an wrongly-mapped neighbor to get a high score. To solve this problem, we work in an iterative manner. In each iteration we select the most possible actual wrong mappings to correct, and recalculate scores of all the related neighbors to find the to-be-corrected mappings in the next turn. 

\begin{figure}
    \centering
    \includegraphics[width=\linewidth]{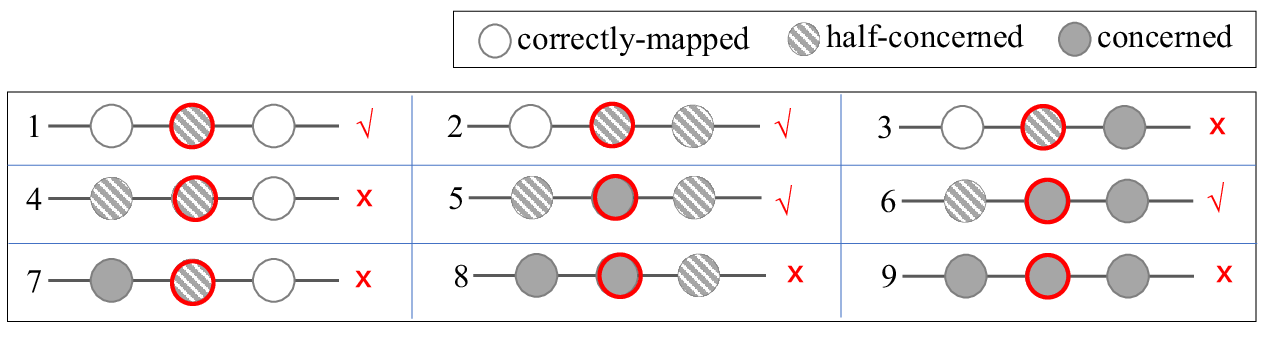}
    \caption{9 situations a wrongly-mapped IP could have in a traceroute path.}
    \label{fig_ip_situation_in_ml}
\end{figure}

To select wrong mappings to be corrected now, we consider 9 situations a wrong mapping could have in a traceroute path as depicted in Fig.\ref{fig_ip_situation_in_ml}. We classify wrongly-mapped IPs into \emph{half-concerned}, if the IP has at least one correctly-mapped neighbor, and \emph{concerned} otherwise. We consider \emph{concerned} IPs to have higher correction priority than \emph{half-concerned} IPs, while correctly-mapped IPs have the lowest correction priority. Then, if a wrongly-mapped IP is in the middle of two neighbors with lower correction priorities (cases $1$ and $5$ in Fig.\ref{fig_ip_situation_in_ml}), we think the mappings of the IP's neighbors are convincing enough for it to make a correction now (in case $5$, the \emph{half-concerned} state of its neighbors might be caused by the wrong mapping of the IP itself, which we will verify in the subsequent turns). Otherwise, if a wronly-mapped IP has one neighbor with the same correction priority and another neighbor with a higher (cases $2$, $4$, $6$ and $8$), we prefer to correct the IP mapping instantly only if the IP is on the left side of its equal-priority neighbor (cases $2$ and $6$), otherwise (case $4$ and $8$) we leave the mapping unchanged in this turn. This is because that, our D-paths are comprised of a few VPs tracerouting to all routable IPv4 prefixes, and thus a transit IP would have much more successors than predecessors. Then, an wrongly mapped successor would have fewer impacts on the scoring of the given IP compared with its predecessor. So we correct the predecessor first and then check if the score of its successors will change. Next, if a wronly-mapped IP has a neighbor with lower correction priority and another neighbor with a higher (case $3$ and $7$), we will leave the IP but correct the neighbor with higher priority in this turn, since the latter is more likely the cause of the two consecutive wrongly-mapped IPs. The last case is when a \emph{concerned} IP is in the middle of two \emph{concerned} neighbors (case 9), and we choose to correct its predecessor first with the same reason as described in cases $2$ and $6$.

We make corrections in a conservative way, i.e., only when a wrongly-mapped IP is decided to be corrected instantly in all of the traceroute paths it appears in, do we launch its correction process. After correcting all the selected IPs in one turn, we recalculate the mapping scores of all the related neighbors and start the next turn. The iteration terminates when less than $N$ (we set $N$ to 50) IPs can be corrected in one turn.

\subsection{Evaluation} \label{sec_eval_ml}
We call the whole process, including constructing the model and correcting wrong mappings, as LearnToCorrect. We evaluate LearnToCorrect in two aspects. First, we calculate LearnToCorrect's mapping error ratios using the ground truth data in \$\ref{sec_eval_IP_gt} and VISV data in \$\ref{sec_visv}. Second, we use LearnToCorrect to re-perform our CD comparison on all the datasets, and compare it with the other IP-to-AS mapping methods with respect to trace-level mismatch ratios. 

Table \ref{tab_eval_gt} lists that LearnToCorrect has an average mapping error ratios of 0.5\% and 0.8\% respectively using the Intra-AS and ISP ground truth, which is apparently lower than all the existing mapping methods. It also achieves an average error ratios of 1.4\% evaluated by VISV, which means that it corrects 70\% of BdrmapIt's mapping errors. The CDF of LearnToCorrect's mapping error ratios is also depicted in Fig.\ref{fig_eval_visv} (green rectangular). Note that LearnToCorrect can map out nearly all the IPs that BdrmapIt cannot map without bringing extra errors, which account for 6,000$\sim$7,000 transit IPs. 

Fig.\ref{fig_trace_level_mismatch} shows that LearnToCorrect (green squares) achieves the lowest mismatch ratios of 11\% on average. It indicates that LearnToCorrect can correct BdrmapIt's mapping errors on ``critical'' IPs, while the overall enhancing mapping accuracy helps to reduce mismatched cases caused by mapping errors.

\section{Study Mismatch Cases} \label{sec_study_mm_cases}
\subsection{Identify possible real-mismatched pairs}
Although LearnToCorrect achieves the lowest mapping error ratios and lowest trace-level mismatch ratios, there may still be errors in the mapping results that cause CD pairs mismatch. We filter possible mapping errors first before studying the real-mismatched cases.

Previous studies assume that a mismatched segment pair with no more than two mismatched AS hops is caused by mapping errors~\cite{ref_jsac}; or more strictly, a single mismatched AS hop with the corresponding C-path and D-path hops being neighbors or siblings are caused by mapping errors~\cite{ref_sig_mao}. However, these methods are aggressive in identifying mapping errors and can fail even in simple real-mismatched cases. For example, AS $A$ announces to directly peer with AS $B$, while its router forwards packets along the default route to $A$'s provider $C$ who also peers with $B$, and $C$ then forwards the packets to $B$. In this case, the mismatched hop $C$ would be considered as a mapping error by both of the above methods.

We try to identify real-mismatched cases as many as possible. For a mismatched segment pair $S$, we consider it a real mismatch if it contains more than two mismatched hops (the same as \cite{ref_jsac}), or if it contains a mismatched hop $H$ that fulfills one of the following conditions: (1) the mapping of $H$ results in more matched segment pairs than mismatched segment pairs, which is the same rationale as in VISV (\$\ref{sec_visv}); (2) there are other neighboring IPs with different address blocks (we check address blocks instead of IPs to exclude cases when a same address block shared between neighbors or siblings may be mapped to a wrong AS) being mapped to the same AS as $H$, which indicates that packets are highly-likely forwarded through that AS rather than multiple concurrent mapping errors. In a word, we extend the considerations in \cite{ref_jsac}, and strive to identify real mismatches even when a mismatched segment pair contains no more than two mismatched hops\footnote{We do not take the assumptions in \cite{ref_sig_mao} since the forwarding policy in reality is too complex to conclude that the CD correspondents having a relationship is a mapping error.}. Our method utilizes the comparison results of massive path pairs (condition 1) and checks the status of the IP-level hops instead of the coarser AS-level hops (condition 2). 

\begin{figure}
        \includegraphics[width=\linewidth]{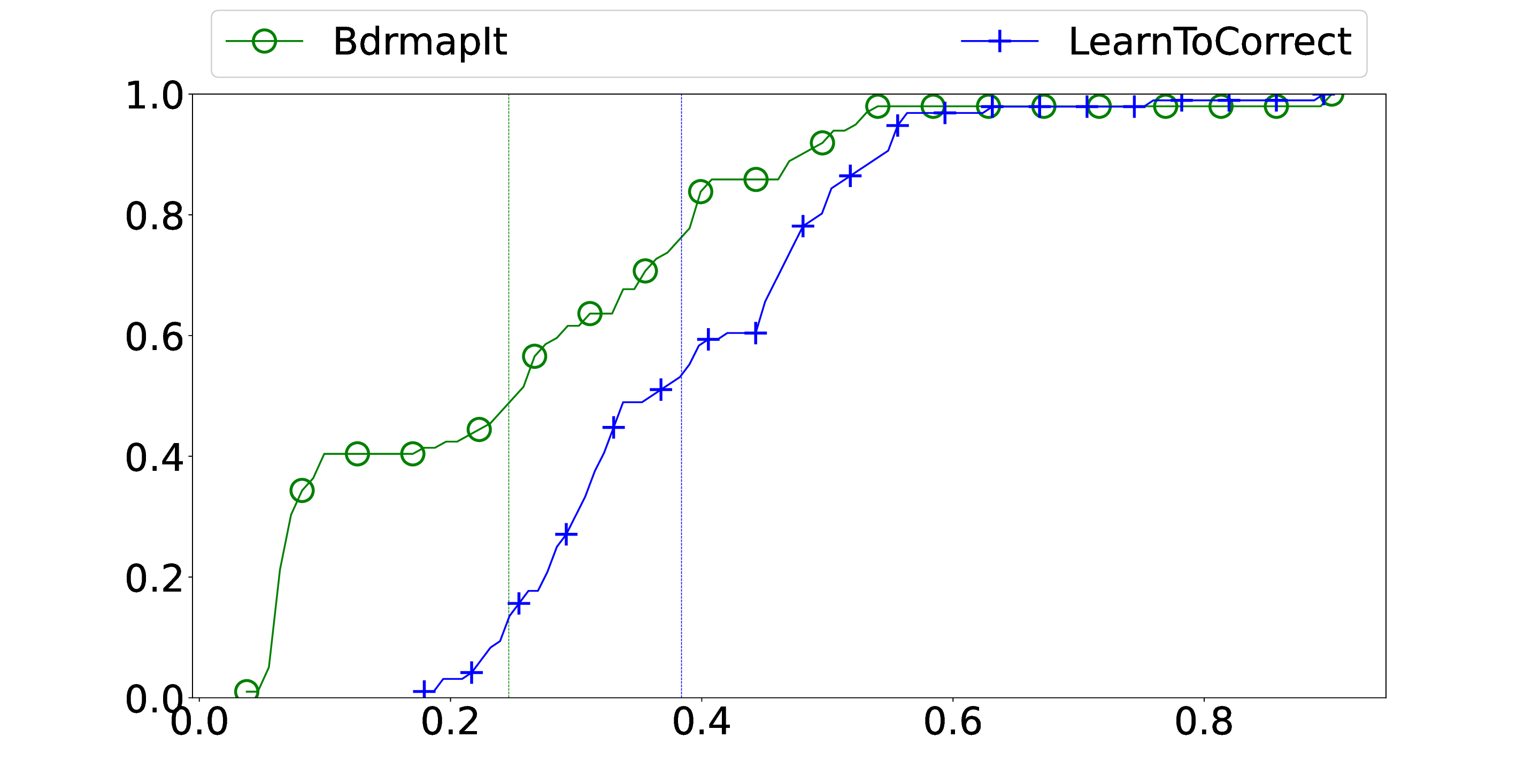}
  \caption{CDFs of real-mismatch ratios among the mismatched pairs.}\label{fig_rrm}
\end{figure}

We perform the above procedure respectively to the mismatched path pairs found by BdrmapIt and by LearnToCorrect. Fig.\ref{fig_rrm} shows the CDFs of the ratios of the identified real-mismatched path pairs among the mismatched pairs under each mapping method. 
We can see that less than 22\% mismatched path pairs under BdrmapIt (blue curve) has been identified as real mismatches, which is lower than the 39\% under LearnToCorrect (orange curve). The results demonstrate from another side that the mismatched pairs under BdrmapIt contain more mapping errors than those under LearnToCorrect. 

Given the high mapping accuracy of LearnToCorrect, we then study the real-mismatched pairs identified under LearnToCorrect, which on average account for 6\% of all the path pairs in our datasets. 

\begin{figure}
  \centering
  \includegraphics[width=.95\linewidth]{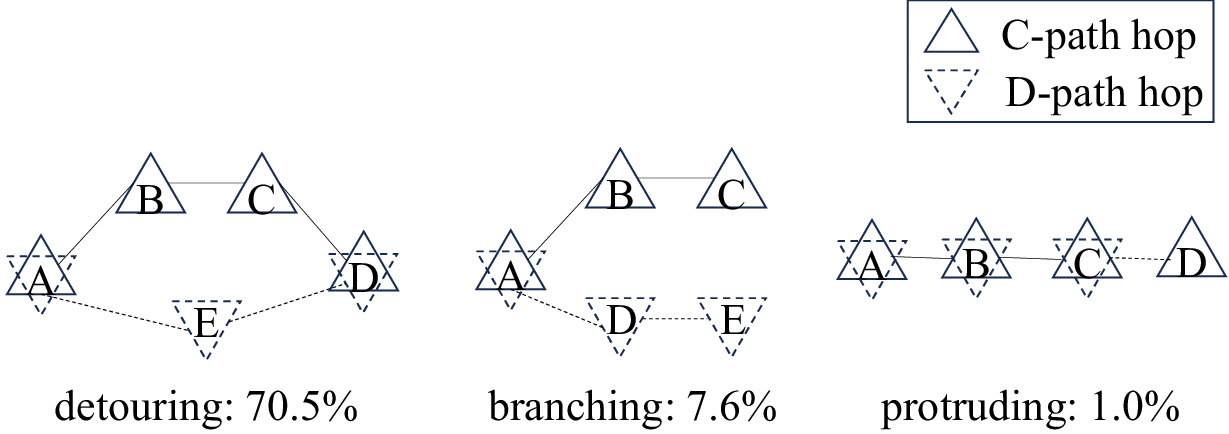}
  \caption{A pattern of CD mismatch in reality.} \label{fig_rm_types}
\end{figure}

We categorize the mismatched cases into 3 types: \emph{detouring}, \emph{branching} and \emph{protruding}. As depicted in Fig.\ref{fig_rm_types}, \emph{detouring} refers to that the D-path differentiates with the C-path at some hop but converges to the C-path at a further hop; \emph{branching} refers to the cases that the D-path differentiates with the C-path at some hop, and then extends toward another direction without returning back to the C-path; \emph{protruding} refers to that the D-path matches the whole C-path except for some extra tails. Each type accounts for 70.5\%, 7.6\% and 1.0\% cases respectively. 

As stated before, CD mismatch can be caused by various legitimate reasons. One common scenario is route aggregation that results in CD mismatch with the pattern of \emph{protruding}. However, some mismatched cases may indicate BGP security issues. For example, a specific prefix hijacking (hidden hijack, as explained in \$\ref{sec_hh}) can result in CD mismatch with the pattern of \emph{branching}; and bogus BGP link (i.e., path hijacking~\cite{ref_mitm}) can result in CD mismatch with the pattern of \emph{detouring}. In the following two subsections, we show how to utilize real-mismatched cases to find out these two kinds of hijackings that are usually considered hard to detect in reality.

\subsection{Hidden Hijack}\label{sec_hh}

\begin{figure}[!h]
  \centering
  \includegraphics[width=\linewidth]{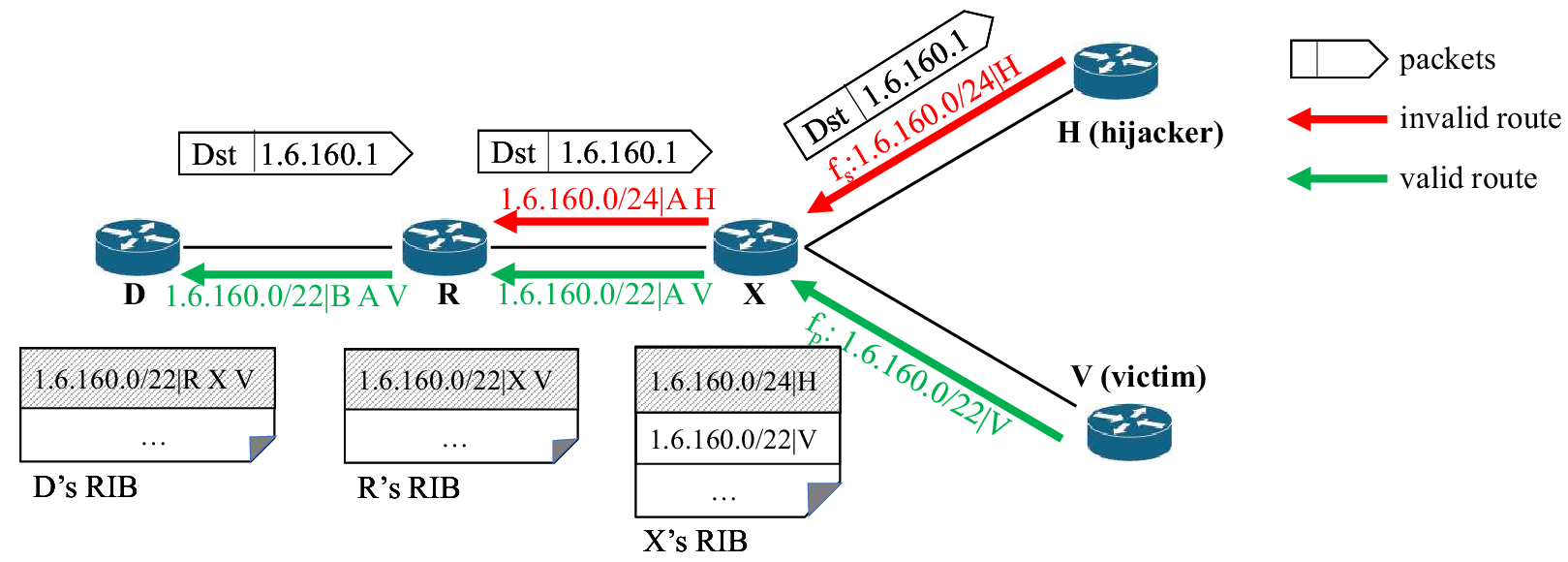}
  \caption{Process of \emph{H} launching a hidden hijack.}
  \label{fig-hh}
\end{figure}

Hidden hijack (HH) is a sub-prefix BGP hijack where in the control-plane a router does not see BGP routes 
indicating the hijack, but in the data-plane, its traffic towards that sub-prefix is actually hijacked. Traditionally, simple prefix hijacking happens when a 
 hijacker AS $H$ announces a prefix $f$ not owned by itself via BGP to attract traffic, 
which has long been recognized as a foundational security flaw of BGP \cite{feamster2004some, 
ref_argus}. In recent years, the Resource Public Key Infrastructure (RPKI)~\cite{ref_rpki} has been proposed to protect prefixes from been hijacked. 
An AS can certify its legitimate ownership of prefixes with Route Origin Authorization (ROA)
. When some AS $R$ receives a BGP message for a prefix $f$ which has ROA records, 
$R$ can verify 
whether $f$ should be originated by the BGP ORIGIN-AS in the message, and choose to discard and not propagate the message if this Route Origin Validation (ROV) procedure fails.
In this context, hidden hijack arises when ROV is only partially deployed in the Internet \cite{ref_rov++}\footnote{The adoption rate of ROV is estimated to reach only around 4\%, and seems to be growing only at a sluggish pace \cite{ref_rov_mi, ref_apnicrov}.}.  Fig.\ref{fig-hh} depicts such an example: an invalid announcement for a sub-prefix $f_s$ of some parent prefix $f_p$ can be dropped by $R$ with ROV, and a downstream AS $D$ will not receive the invalid announcement.  When $D$ forwards traffic destined to $f_s$, $D$ would forward the traffic to an AS $X$ according to its BGP route for $f_p$. If $X$ has not deployed ROV and is hijacked by the illegitimate announcement of sub-prefix $f_s$,  $X$ will then forward the traffic to the illegitimate ORIGIN-AS $H$. 

Different from traditional prefix hijacking where the control-plane and data-plane match, HH results in a CD mismatch. Thus, we examine real-mismatch cases where (1) the D-path does not reach or pass through the BGP ORIGIN-AS $V$; (2) some other VP can see the destination AS $H$ on the traceroute path announcing a subprefix $f_s$ of the parent prefix $f_p$, which covers 
the destination IP; (3) $V$ has registered ROA for prefix $f_p$ but $H$ has not for $f_s$; and (4) $V$ and $H$ are not siblings. The last step is to filter cases of legal address reuse within an organization.

We apply this method to the datasets of 2022.12.15 which have 2.4M path pairs, and 260K among them are mismatched under LearnToCorrect. After we apply the above conditions, we gradually have 12K, 906, 52 and 20 pairs left. The final 20 possible HH pairs relate to 18 $<$prefix $f_p$, victim AS $V$, Hijacker AS$H >$ tuples. Note that our observation is based on a very limited scale (3 Ark VPs and several Atlas VPs on that date, with each Atlas VP tracerouting to only 100 IPs). Given that the current partial deployment of ROVs leaves ample room to launch hidden hijacks, a specific method to detect HH on the international scale is needed, and our initial attempt shows how CD mismatch can be used in the detection. Moreover, since hidden hijacks often occur when a high-tier AS deploys ROV while its customers does not, the control-plane and data-plane path may differ mostly in a few last hops. Thus, an accurate mapping method, e.g., LearnToCorrect, is crucial to achieve the detection.  

\subsection{Bogus link}
Besides prefix hijacking, an AS can also manipulate the BGP paths in BGP announcements. A typical scenario is to forge a bogus ``shortcut'' BGP link that does not exist, to make the announced BGP path shorter and more attractive. This kind of path hijacking cannot be detected by the currect RPKI system. Especially, bogus links can be used in Man-In-The-Middle(MITM) intreception attacks, where hijackers manage to forward the traffic to the legal origin AS after successfully attracting the traffic, which makes the attacks even harder to detect\cite{ref_mitm, ref_interception, ref_argus, ref_artemis, ref_spurious_bgp}. 

However, bogus links in interception trigger CD mismatch with a special mismatch pattern which we call \emph{link-detour}. In a \emph{link-detour}, the C-segment is simply a link 
$AS_x$--$AS_y$, while the mismatched D-segment contains additional hops in the middle, e.g., 
$AS_x$--$AS_1$--$AS_y$.
We still use our mismatched pairs found in 2022.12.15, and find 443 \emph{link-detour} cases. We then validate the \emph{suspicious} BGP links in these cases in both passive and active ways. 

\textbf{Passive filtering.} First, we download all the traceroute paths from Ark and Atlas platforms on 2022.12.15. We use LearnToCorrect to map the IPs in the traceroute and get a dataset of over 544,000 AS-level data-plane links. 332 \emph{suspicious} links can be found in the dataset and are thus filtered. 

\textbf{Active filtering.} For the remaining 212 links, we lauch active traceroute using Atlas VPs (i.e., probes). If an Atlas probe can be found within either endpoint AS of a \emph{suspicious} link, we use that probe to perform traceroute to the live IPs in the other endpoint AS, and check if the concerned link is used in data-plane. 97 links can be found in the above measurements. If none of the endpoint AS of a \emph{suspicious} link has an Atlas probe, we collect the BGP paths announced on that day that contains the link, and find if any ASes on the paths have Atlas probes. We perform traceroute from all satisfying ASes respectively to the farther endpoint ASes of the link to check the link's appearance. Finally, 70 \emph{suspicious} links are filtered out in this way. 

For the remaining 45 links that are absent from any data-plane paths we can gather, we manually validate them by contacting the network operators of the endpoint ASes. We received 7 valid responses, where 4 links do not exist (3 due to expired usages and 1 due to unknown reason), 1 link is a GRE tunnel provided by a tier-1 AS (i.e., only logically exists), while 2 links actually exist. 

In this experiment, we use CD comparison to filter out 89.9\% \emph{suspicious} links and get suspicious links with 71\% accuracy. Our experiment exemplifies how to use CD mismatch to find bogus links, while a more systematic detection method can be studied in future. Again, an accurate mapping method is essential in the comparison process.

\section{Conclusion}
In this paper, we conduct a large-scale and long-term measurement on the mismatch extent of control-plane and data-plane paths in the Internet. Although this mismatch is allowed by design of the Internet, it's pervasiveness and impact on the Internet is still far from clear. We study data collected from 128 pairs of VPs over more than 5 years, and use 6 state-of-art mapping methods to perform the CD comparison. The trace-level mismatch ratios vary sharply from 8\% to 69\% on different datasets under different mapping methods. To study the efficiency of the mapping methods, we propose VISV to generate large-scale accurate mappings to provide as semi-ground truth. We show that the BdrmapIt based methods outperforms the traditional RIB based methods with apparently higher accuracy, while they may wrongly map ``critical'' IPs that occurs frequently in traceroute paths, which then greatly lower the trace-level mismatch ratios. We propose a new mapping method LearnToCorrect that can achieve both the lowest mapping error ratios and the trace-level mismatch ratios. We devise to identify real-mismatch cases from mismatched pairs, and shows how to use them to detect routing security issues. We release all our codes and results to facilitate future studies. 

\printbibliography

\begin{appendices}


\section{Bifurcate at the second AS hop}\label{append-first-bifurcate}
In our original datasets, there are path pairs which bifurcate right after the source AS after IP-to-AS mapping under every mapping method. 
We call this mismatch pattern as \emph{FirstBifurcate}.

When the C-VP and D-VP have the same routing policy, \emph{FirstBifurcate} can happen if the D-VP configures packets redirection rules, or has a partial FIB~\cite{ref_forwarding_detours} which makes packets to be forwarded through a default route and to leave the source AS with an exit different from what its control-plane tells.

However, since we conduct a 5-year-comparison study, either C-VP or D-VP within a pair would change its routing or forwarding policy during the period. Moreover, 
some C-VPs choose not to announce some routes to the route collectors their peer with. Both cases would result in the form of \emph{FirstBifurcate} and make CD comparison inaccurate. In this section we compare the original path pairs (i.e., do not filter \emph{FirstBifurcate}) using our mapping method LearnToCorrect, and distinguish \emph{real-mismatch} cases using IASC. Based on which we discuss representative cases of \emph{FirstBifurcate}.

\subsection{Policy change (1): a burst of \emph{FirstBifurcate} in Ark D-VP \emph{ams-nl}}
\begin{figure}
  \includegraphics[clip,width=\linewidth]{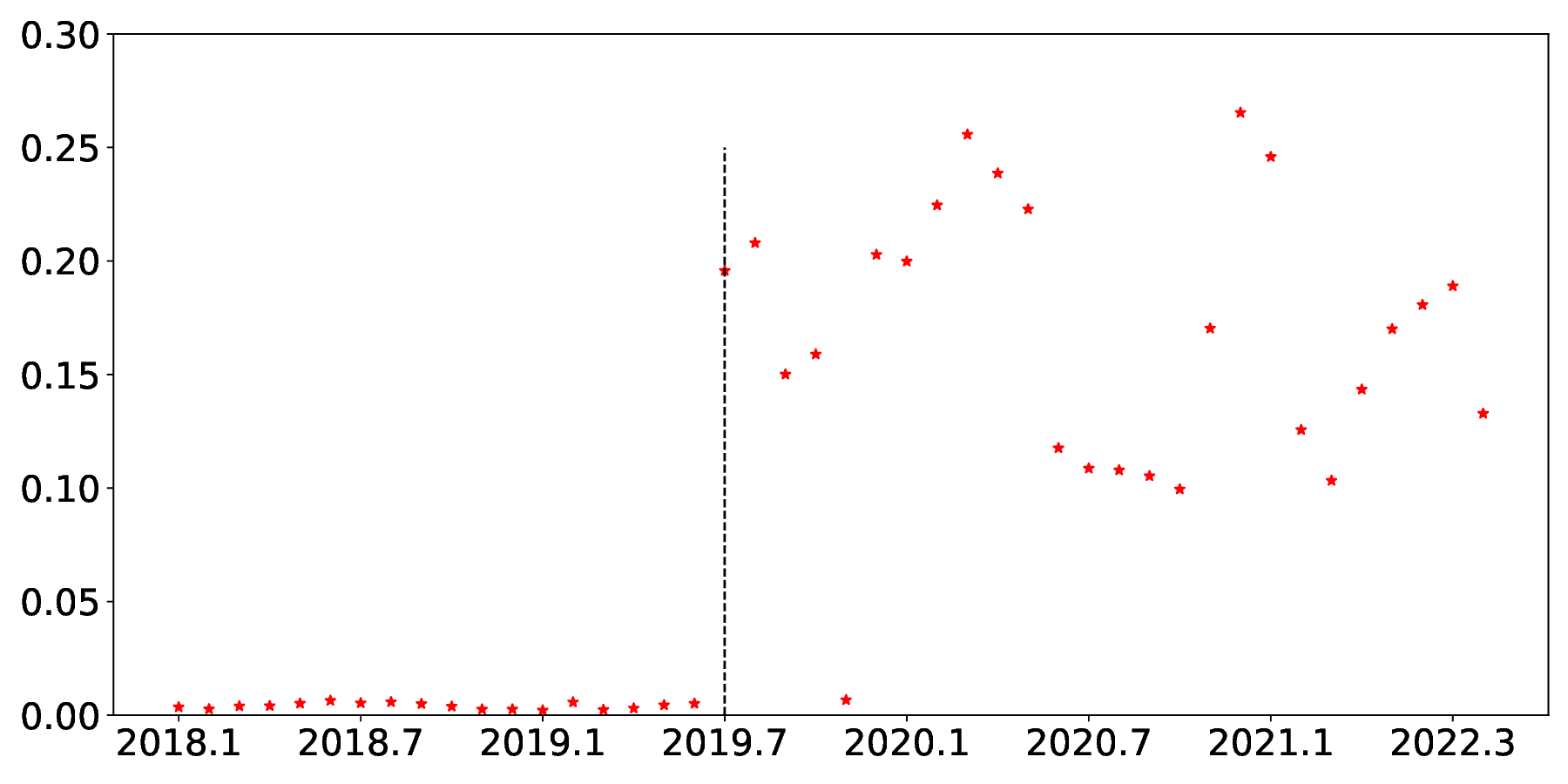} 
  \caption{Absolute \emph{r-m} ratios of \emph{ams-nl} along time.}
  \label{fig_as1103_rm}
\end{figure}

Fig.~\ref{fig_as1103_rm} shows the absolute \emph{r-m} ratios of the path pairs (including \emph{FirstBifurcate} cases) of Ark D-VP \emph{ams-nl} during 2018.01$\sim$2022.04 (the path pairs only provide data on 7 days since 2021.01). We can see that the ratios rise sharply since 2019.07. A first look at the AS-level shows that, in over 68\% mismatched path pairs, the second AS hop in the D-path, i.e., AS3257 (sometimes AS286), differs from the second hop in the C-path, i.e., AS286 (sometimes AS3257). Learning that AS286 began to merge to AS3257 since 2019.07, it looks as if these mismatched cases are caused by mapping errors (address reuse among sibling ASes). However, a further look shows that it is not the case. Using TSA to annotate mapping status, these second-hop mappings are annotated as \emph{other}, denoting that they can also make many path pairs match.

\begin{figure}
  \includegraphics[clip,width=\linewidth]{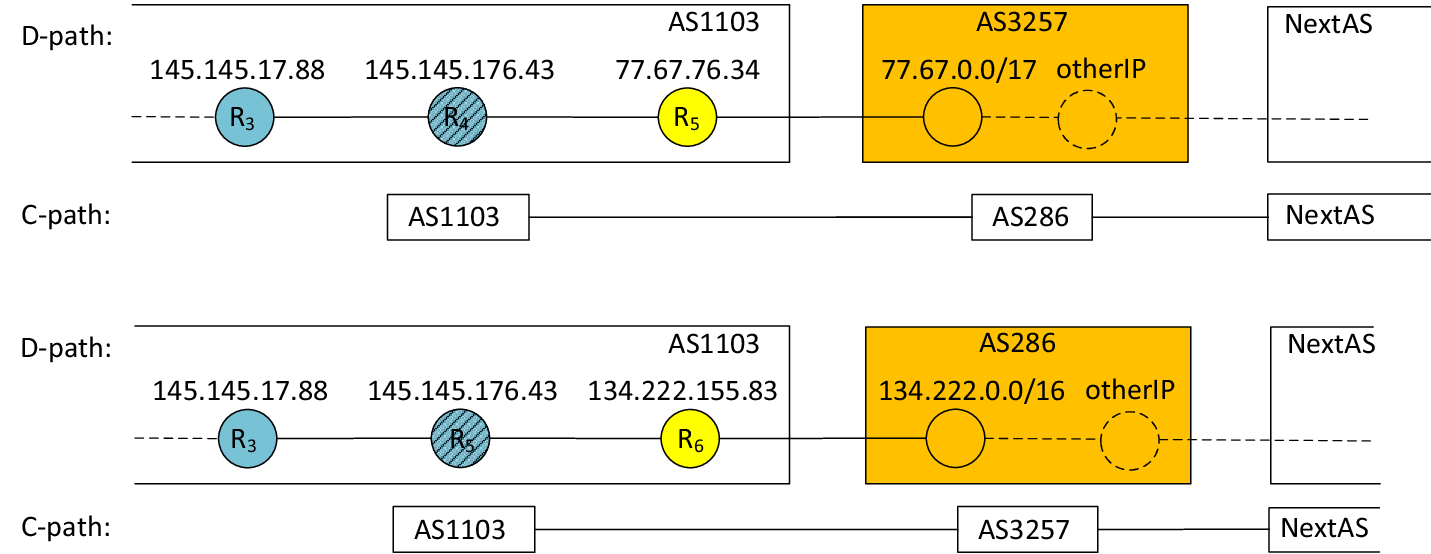} 
  \caption{Typical mismatch patterns at \emph{ams-nl} since 2019.07.}
  \label{fig_as1103_set2}
\end{figure}

We check on the IP-level traceroute paths and find that, before 2019.07, all traceroutes launched from {\itshape ams-nl} go through the same first three hops.
After the third hop 145.145.17.88 (the blue circle labeled $R_3$ in Fig.\ref{fig_as1103_set2}), 
no single forth hop appears in more than 30\% of all the paths. However, things changed since 2019.07, where $R_3$ chose 145.145.176.43 (the blue shaded circle labeled $R_4$) as its next-hop in over 98\% traceroute paths. This default forth hop $R_4$ sometimes chose to forward packets to AS286/AS3257 in the data-plane, while in the control plane the next-hop AS is AS3257/AS286 (as shown in Fig.~\ref{fig_as1103_set2}), which account for 53\% of all the mismatched cases. 

\begin{figure}
  \includegraphics[clip,width=0.95\linewidth]{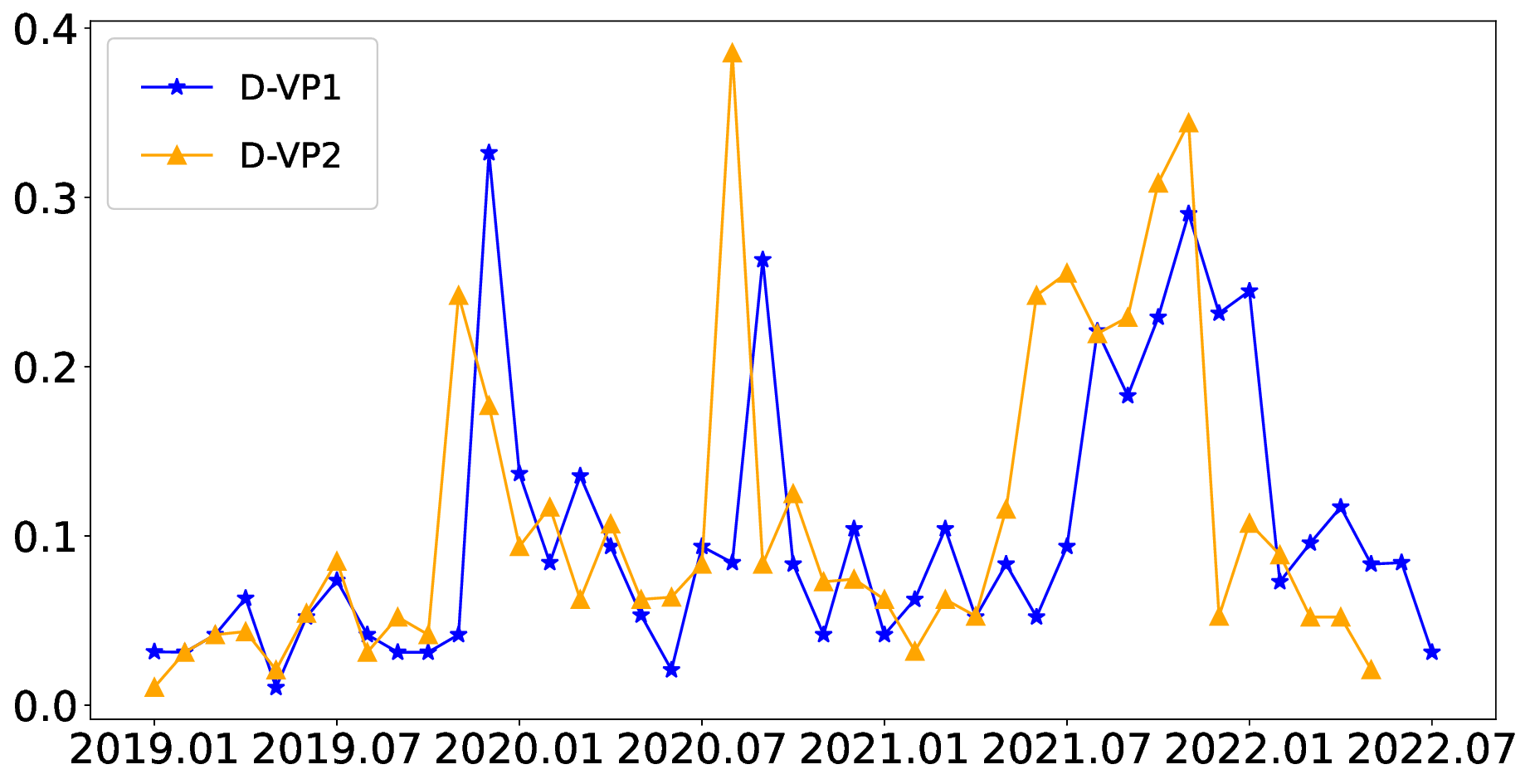} 
  \caption{Absolute \emph{r-m} ratios of the two VP pairs of AS18106.}
  \label{fig_spec_atlas}
\end{figure}

\begin{table*}
\centering
\setlength\tabcolsep{5pt}
\caption{Feature set.}\label{feature-set}
\begin{threeparttable}
\begin{tabular}{ccc}
\toprule
name&importance&description\\
\hline
\emph{bdr-rib-rel}&0.02&relationship of the current mapping and the RIB+PeeringDB mapping\\
\emph{prev-sameAS-rate}&0.04&ratio of the predecessor hops whose mappings are the same as self\\
\emph{succ-sameAS-rate}&0.12&ratio of the successor hops whose mappings are the same as self\\
\emph{succ-ip-uncertain-rate}&0.01&ratio of predecessor hops being unresponsive/\emph{unmap}\\
\emph{prev-asn-uncertain-rate}&0.02&ratio of predecessor ASes having unresponsive/\emph{unmap} hops in between\\
\emph{succ-asn-uncertain-rate}&0.06&ratio of predecessor ASes having unresponsive/\emph{unmap} hops in between\\
\emph{valley-normal-rate}&0.29&ratio of the AS-level triples being valley-free\\
\emph{valley-abnormal-rate}&0.12&ratio of the AS-level  triples being not valley-free\\
\emph{valley-seminormal-rate}&0.16&ratio of the AS-level triples being semi-valley-free\\
\emph{prev-asnrel-unknown-rate}&0.01&ratio of predecessor ASes having no relationship with the self\\
\emph{succ-asnrel-unknown-rate}&0.13&ratio of successor ASes having no relationship with the self\\
\bottomrule
\end{tabular}
    \end{threeparttable}
\end{table*}

In the beginning VP pair selection process, we found two comparable C-VPs for the Ark D-VP \emph{ams-nl}, i.e., 80.249.208.34 (we name it $C_1$) from \emph{rrc03} and 80.249.208.50 ($C_2$) from \emph{route-views.amsix}. We examined the RIBs of the two C-VPs on 2018.01.15 00:00:00, and found that the two RIBs are basically the same. So we randomly picked $C_1$ as the C-VP for \emph{ams-nl}. Given the vast mismatched cases since 2019.07, we re-check the RIBs of $C_1$ and $C_2$ in 2019.07, and find that over half of the routes differ. The traceroute paths in \emph{ams-nl}, meanwhile, can match around 80\% C-paths in $C_2$ under the CorrectToLearn mapping method.

All the above observations suggest that either the D-VP \emph{ams-nl} or the C-VP $C_1$ changed their routing or forwarding policy since 2019.07 (perhaps it is related to the merge of AS286 and AS3257, both of which are providers of AS1103 who operates \emph{ams-nl}), and make the path pairs of \emph{ams-nl} and $C_1$ incomparable.

\subsection{Policy change (2): fluctuate \emph{FirstBurcate} cases in Atlas D-VPs of AS18106}
We have two VP pairs that both reside in AS18106. The two VP pairs have the same C-VP but with different D-VPs. 
We check the absolute \emph{r-m} ratios (including \emph{FirstBifurcate} cases) during the experiment period for each VP pair. As shown in Fig.\ref{fig_spec_atlas}, the absolute \emph{r-m} ratios on the two VP pairs fluctuate with similar variation trends. We look into the datasets where the absolute \emph{r-m} ratios are higher than 10\%, and find that in most cases, the source AS18106 chooses one provider (e.g., AS2914) as its next-hop in the control-plane, while another provider (e.g., AS174) as its next-hop in the data-plane.

Given that the destinations that each D-VP traceroutes to are random and different, the highly conformed absolute \emph{r-m} ratios between the two VP pairs suggest that it may be the policy change of the C-VP or the default router of the two D-VPs (many Atlas D-VPs are end-users instead of routers) that leads to a large scale of \emph{FirstBifurcate}.

\subsection {Selective announcements: specific \emph{FirstBifurcate} patterns in Ark D-VP \emph{nrt-jp} and \emph{sao-br}}
Since 2021, in the \emph{real-mismatch} cases (again, including \emph{FirstBifurcate} cases) in Ark D-VP \emph{nrt-jp}, 55\% cases have the following pattern:
\begin{center}C-path: AS7660, $next$-$hop$, ...\end{center}
\begin{center}D-path: AS7660,  AS6939,  ...\end{center}
In these cases, the source AS7660 has the next-hop of one provider (mostly AS2516) in the control-plane, while choose to go through AS6939 in the data-plane. Nearly all the prefixes we found in the C-VP's rib for the traceroute destinations in these cases have the prefix length less than 24. Moreover, almost for each destination in the cases, we can find from another C-VPs a route to its "/24" prefix announced by AS6939, which conforms the mismatched D-segment in our path pairs. We contact the network operator of AS7660 who confirms that AS7660 chooses not to advertise routes it learns from JPIX TOKYO, according to its local policies. Thus, the incomplete RIBs of the C-VP make us find a route to a less-specific prefix which then mismatches the data-plane.

Similar cases happen at Ark D-VP \emph{sao-br}. The operator of AS22548 (who operates \emph{sao-br}) told us that the C-VP we use only announces to its route collector the peering routes learned from IX-São Paulo and from its private peerings. The incomplete RIBs of the C-VP cause many mismatched cases, mainly \emph{FirstBifurcate}.

From the above case studies we see how incomparable VP pairs or incomplete RIBs cause a large amount of \emph{FirstBifurcate} cases. Since we cannot distinguish them from those configuration issues within the D-VP (packet redirection or partial FIB etc), and more importantly, we concern the inter-domain CD mismatch that may cause security concerns, we discard the \emph{FirstBifurcate} cases from our study.

\section{Features of the model in LearnToCorrect}\label{appendix_ml}
Table \ref{feature-set} lists all the features we use in classifying IP mappings in the machine learning process. 

\end{appendices}
\end{document}